\documentclass[aps,prb,twocolumn,superscriptaddress,nofootinbib]{revtex4-2}
\usepackage{amssymb,graphicx,color}
\usepackage[intlimits]{amsmath}
\usepackage[english]{babel}
\usepackage[colorlinks]{hyperref}
\hypersetup{
	colorlinks=blue,
	linkcolor=blue,
	filecolor=blue,
	urlcolor=blue,
	citecolor=blue
}
\usepackage[normalem]{ulem}
\usepackage{comment}
\usepackage{ragged2e}
\usepackage{enumerate}
\usepackage{subfigure}
\usepackage{float}
\usepackage{ulem}

\begin{document}

\title{Phase transitions of the anisotropic Dicke model }

\author{Pragna Das}
\affiliation{Indian Institute of Science Education and Research Bhopal 462066 India}
\author{Devendra Singh Bhakuni}
\affiliation{Department of Physics, Ben-Gurion University of the Negev, Beer-Sheva 84105, Israel}
\author{Auditya Sharma}
\affiliation{Indian Institute of Science Education and Research Bhopal 462066 India}

\begin{abstract}
  We systematically analyze the various phase transitions of the
  anisotropic Dicke model that is endowed with both rotating and
  counter-rotating light-matter couplings. In addition to the ground
  state quantum phase transition (QPT) from the normal to the
  super-radiant phase, the anisotropic Dicke model also exhibits other
  transitions namely the excited state quantum phase transition
  (ESQPT), ergodic to non-ergodic transition (ENET) and the
  temperature dependent phase transition. We show that these phase
  transitions are profitably studied not only with the standard
  consecutive level spacing ratio, but also with the aid of various
  eigenvector quantities such as von Neumann entanglement entropy, the
  participation ratio, multifractal dimension and mutual
  information. For ENET, both the statics and dynamics of the
  participation ratio offer a consistent and useful picture. An
  exciting finding from our work is that the ESQPT and the ENET are
  closely related to each other. We show this with the aid of two
  characteristic energies in the spectrum corresponding to jumps in
  von Neumann entropy.
\end{abstract}

\maketitle 

\section{Introduction}\label{sec_1}

The Dicke model~ \cite{dicke1954coherence, lambert2004entanglement,
  emary2003chaos, emary2003quantum}, which is paradigmatic within the
field of cavity quantum electrodynamics, describes the interaction
between $N$ atoms and a single-mode bosonic field via a dipole
coupling strength. In the thermodynamic limit ($N\to\infty$), the
model shows a quantum phase transition from the normal phase (NP) to
the super-radiant phase (SP)~\cite{dicke1954coherence,
  kadantseva1990superradiance, chavez2016classical,
  kirton2018superradiant, kirton2019introduction,
  prasad2022dissipative, das2022revisiting} at some critical coupling
strength. Along with this quantum phase transition (QPT), the Dicke
model also exhibits two other distinct phase
transitions~\cite{yamamoto2020quantum}, namely the excited state
quantum phase transition (ESQPT)~\cite{brandes2013excited,
  bastarrachea2014comparative, chavez2016classical,
  stransky2016classification, stransky2017excited, cejnar2021excited}
and the thermal phase transition (TPT)~\cite{carmichael1973higher,
  wang1973phase, duncan1974effect, Bastarrachea-Magnani_2017}.  While
the first occurs at finite energy when the coupling strength is
sufficiently large, the second, on the other hand, occurs at a finite
temperature~\cite{wang1973phase}. Some of these quantum phase
transitions were observed experimentally in Bose-Einstein
condensates~\cite{rubeni2012quantum} and quantum cavity
systems~\cite{greentree2006quantum}.  The physics of systems with
light-matter interactions has enjoyed a great deal of interest in
recent times, triggered by a number of experimental
works~\cite{stranius2018selective, zhou2019emerging, mueller2020deep}.
        
A generalized version of the Dicke model, namely the anisotropic Dicke
model~\cite{hioe1973phase, de1991classical, de1992chaos,
  furuya1992husimi, bastarrachea2016thermal,
  buijsman2017nonergodicity, kloc2017quantum, aedo2018analog,
  shapiro2020universal, cejnar2021excited, hu2021out} (ADM), where the
coupling strengths corresponding to the rotating and counter-rotating
terms are different, has gained traction in recent times. While a huge
body of literature has been built around the Dicke
model~\cite{baumann2011exploring, bhattacharya2014exploring,
  perez2017thermal, ashhab2019spectrum, lewis2019unifying,
  kirton2019introduction}, the anisotropic model has received
relatively less attention. The asymmetry in the coupling brings some
novel features in addition to the existing properties of the Dicke
model. One such novel feature that has generated considerable
excitement is that the ADM not only exhibits the normal to
super-radiant quantum phase transition, but also an
ergodic-to-nonergodic transition
(ENET)~\cite{buijsman2017nonergodicity, hu2021out}. The model is
integrable in the limit where either one of the couplings is zero.
Moreover, while the ground state properties show the
normal-to-super-radiant phase transition, the excited states show
signatures of non-ergodicity~\cite{hu2021out}.  It was also argued
that the transition from the normal to the super-radiant phase is
quite different in comparison to the ergodic to non-ergodic
transition~\cite{buijsman2017nonergodicity}. In the
  present work, focusing on eigenvector properties, we show that the
  normal-to-super-radiant phase transition corresponds to the ground
  state undergoing a localized-to-multifractal transition. On the
  other hand the ergodic-to-nonergodic transition corresponds to
  the middle excited state undergoing a delocalized-to-multifractal
  transition.

Phase transitions are often characterized by quantum information tools
such as entanglement entropy~ \cite{lambert2004entanglement,
  bardarson2012unbounded, bera2015many, roy2018entanglement,
  nehra2018many}, mutual information~\cite{divincenzo2004locking,
  lu2011optimal, henderson2001classical, adesso2010quantum} and so on. These
quantities have proven useful not only to mark a variety of phase
transitions ~\cite {bardarson2012unbounded, roosz2014nonequilibrium,
  bera2015many, roy2018entanglement, nehra2018many,
  sirker2014boundary, cho2017quantum}, but in diverse other
contexts~\cite{ma2013multipartite,sharma2015landauer,
  sable2018landauer} where quantum correlations have an important
role.  Moreover, some of these quantities are directly related to
experimentally measurable observables and have proven to be useful
markers of the phase transitions of the Dicke
model~\cite{lewis2019unifying}. Thus, a study of these quantities in
the context of phase transitions is of theoretical interest with
potential to connect with experimental work.
    
  In this work, we explore the various phase transitions of the
  anisotropic Dicke model and their dependence on the asymmetric
  coupling strengths. First, we study the behavior of the well known
  quantum phase transition of the ADM using quantum information
  measures. With the aid of the ground state energy, average photon
  number, inverse participation ratio and its scaling with the
  Hilbert-space dimension, we highlight that the normal to
  super-radiant phase transition is reminiscent of a localization to
  multifractal phase transition. Next, we highlight the emergence of
  the excited state quantum phase transition and the temperature
  dependent phase transition and their dependence on the coupling
  parameters. While these transitions have been studied extensively
  for the Dicke model~\cite{das2022revisiting, perez2017thermal,
    hepp1973superradiant, wang1973phase, brandes2013excited,
    lewis2019unifying}, they are also prominently present in the
  ADM~\cite{buijsman2017nonergodicity, hu2021out}. We study the
  temperature-dependent phase transition as a function of the rotating
  and counter-rotating coupling strengths with the aid of an old
  analytical result~\cite{hioe1973phase} for the transition
  temperature. Similar to the Dicke model~\cite{das2022revisiting}, we
  find that mutual information between two spins offers a clear
  signature of the thermal phase transition, which is benchmarked
  against the analytical expression for the critical temperature
  which has already been worked out in the
    literature~\cite{hioe1973phase, bastarrachea2016thermal}.

  Our main result is to show that the ESQPT is profitably studied with
  the help of von Neumann entanglement entropy between the bosons and
  the spins, the average level spacing ratio and two characteristic
  energies that define a central band in the super-radiant phase. With
  the help of participation ratio and multifractal dimension, we show
  that the middle excited state exhibits multifractal behavior in the
  nonergodic phase. Thus the middle excited state behaves in stark
  contrast to the ground state which shows a change from localized to
  multifractal behavior as one goes from the normal to the
  super-radiant phase. The correspondence between the multifractal
  nature of the middle excited state and the nonergodic phase is also
  captured dynamically when we study the participation ratio in a quench
  dynamical protocol. Another exciting finding of our work is that the
  excited state quantum phase transition and the non-ergodic to
  ergodic transition are closely
  related. Specifically, we find that the phase diagram obtained by
  keeping track of the size of the jumps in von Neumann entropy for
  the different eigenstates of the system (which carry signatures of
  the ESQPT), closely resembles the ENET phase diagram.
  While the ESQPT in the anisotropic Dicke model was
    explored in Ref.~\cite{cejnar2021excited}, it is mainly
    focused on the properties of the
    eigenvalues. The authors of Ref.~\cite{kloc2017quantum} studied the atom-field
    and atom-atom entanglement in the anisotropic Dicke model in which
    the couplings are restricted to $g_1\geq g_2$. Our work considers
    a more general parameter regime with arbitrary non-negative $g_1$
    and $g_2$ like in Buijsman $et$
    $al$~\cite{buijsman2017nonergodicity}.
   
The organization of the paper is as follows. In the 
Sec.~\ref{sec_2}, we introduce the model Hamiltonian. Next we
discuss the various phase transitions (QPT, ESQPT, ENET, TPT)
exhibited by the anisotropic Dicke model and their characterization
via tools from quantum information theory. While the QPT, ESQPT, and
ENET are covered in Sec.~\ref{sec_3}, Sec.~\ref{sec_4} is
dedicated to the TPT. Finally in Sec.~\ref{sec_5} we summarize
the main results.

\section{Model Hamiltonian}\label{sec_2} 
The Hamiltonian consists of a single-mode bosonic field coupled to $N$
atoms with anisotropic couplings of the rotating and counter-rotating
terms:
\begin{eqnarray}
  \mathcal{H} = \omega a^{\dagger}a + \omega_{0}J_{z} + \frac{g_{1}}{\sqrt{2j}}(a^{\dagger}J_{-} + a J_{+}) \nonumber \\ + \frac{g_{2}}{\sqrt{2j}}(a^{\dagger}J_{+} + a J_{-}).
\end{eqnarray}
Here the operators $a$ and $a^{\dagger}$ are bosonic annihilation and
creation operators respectively, and $J_{\pm,z}=\sum_{i=1}^
{2j}\frac{1}{2}\sigma_{\pm,z}^{(i)}$ are angular momentum operators of
a pseudospin with length $j$, composed of $N=2j$ spin- $\frac{1}{2}$
atoms described by Pauli matrices $\sigma_ {\pm,z}^{(i)}$ acting on
site $i$. The commutation relations (in units where $\hbar = 1$)
between the various operators are as follows:
\begin{eqnarray}
  \left[a,a^{\dagger}\right]=1, \left[J_z,J_{\pm}\right]=\pm J_{\pm}, \left[J_+,J_{-}\right]=2J_z.
\end{eqnarray} 

The basis of the full Hilbert space of the system is $\{\vert
n\rangle\otimes\vert j,m\rangle\}$ where $\vert n\rangle$ are the
number states of the field satisfying $a^{\dagger}a\vert
n\rangle=n\vert n\rangle$ and $\vert j,m\rangle$ are the Dicke states
satisfying $J_{\pm}\vert j,m\rangle=\sqrt{j(j+1)-m(m\pm 1)} \vert
j,m\pm 1\rangle$. $\omega$ is the single-mode frequency of the bosonic
field while $\omega_0$ is the level splitting of the atoms. $g_{1}$
and $g_{2}$ are the time independent coupling strengths corresponding
to the rotating and counter-rotating light-matter interaction
terms. In the thermodynamic limit, the system shows a second order
quantum phase transition from normal to super-radiant phase at $g_{1}
+ g_{2}=1$. For $g_{1} + g_{2}<1$, the system is in the normal phase
with $\langle a^{\dagger}a\rangle/j \approx 0$ (the bosonic mode is
microscopically excited) and for $g_{1} + g_{2}>1$, it is in the
super-radiant phase with a positive value of $\langle a^{\dagger}a
\rangle/j$ (bosonic mode of the system is macroscopically
excited). Here the expectation value is calculated with respect to the
ground state of the system Hamiltonian. The ADM possesses a parity
symmetry $[H, \Pi] = 0$ with $\Pi = \exp(i\pi[a^{\dagger} a + J_z +
  j])$ having eigenvalues $\pm 1$~\cite{buijsman2017nonergodicity}.
Here, we restrict ourselves to the $+1$ eigenvalue and an even atom
number $N$ and the symmetric subspace where $j=\frac{N}{2}$. Hence the
$(N+1)$ values that $m$ can take are: ($-\frac{N}{2},\ .\ .\ .\ ,\ 0,\ .\ 
.\ .\,\ \frac{N}{2}$). For our numerics we truncate the boson
number to take the values $n=0,\ 1,\ .\ .\ .\,\ n_{\text{max}}$. Thus the
total Hilbert space dimension for the truncated model is: $N_D = (
n_{\text{max}} + 1 )( N + 1 )$. We have checked that all our numerical
results (for the specified $n_{\text{max}}$) are robust against
further increase in $n_{\text{max}}$, thus the truncation is carried
out in such a way that our numerical results are
reliable. Furthermore, we fix $\omega_{0}=1$ as the
  dimension of energy and the other observables are calculated in
  units of $\omega_{0}$.

In the thermodynamic limit (when the number of atoms $N\to\infty$) the
model is analytically solvable using the Holstein-Primakoff
representation~\cite{chang1995generalized, ressayre1975holstein} of
the angular momentum operators $J_z = ( b^{\dagger}b - j )$, $J_+ =
b^{\dagger}\sqrt{2j - b^{\dagger}b}$, $J_- = J_+^{\dagger}$. Here $b$
and $b^{\dagger}$ are bosonic operators that convert the system
Hamiltonian into a two-mode bosonic problem. This allows us to obtain
effective Hamiltonians that are exact in the thermodynamic limit, by
neglecting terms from expansions of the Holstein-Primakoff square
roots~\cite{emary2003chaos}. In the normal phase $g_1+g_2<\sqrt{
  \omega\omega_0}$, the square roots can be expanded directly and the
effective Hamiltonian is
\begin{eqnarray}
  \mathcal{H}^{(1)} = \omega a^{\dagger}a + \omega_0( b^{\dagger}b - j ) + g_1( a^{\dagger}b + ab^{\dagger} ) \nonumber\\ 
  + g_2( a^{\dagger}b^{\dagger} + a b ),
\end{eqnarray}
which is bilinear in the bosonic operators. In this representation, 
the parity operator $\Pi$ becomes $\Pi = \exp\Big( i\pi\Big[ a^{\dagger}a 
  + b^{\dagger}b \Big] \Big)$.
In the super-radiant phase $g_1+g_2>\sqrt{\omega\omega_0}$, both the 
field and the atomic ensemble acquire macroscopic occupations and for 
that one needs to displace the bosonic modes:
\begin{equation*}
  a^{\dagger}\rightarrow c^{\dagger} + \sqrt{\alpha},\hspace{3mm} b^{\dagger}\rightarrow d^{\dagger} - \sqrt{\beta},
\end{equation*}
where the undetermined parameters $\alpha$ and $\beta$ are of order 
$O(j)$. Now considering the thermodynamic limit, the 
Hamiltonian can be written as:
\begin{widetext}
  \begin{eqnarray}
    \mathcal{H}^{(2)} &=& \omega c^{\dagger}c + \left[ \omega_0 + \frac{(g_1+g_2)}{k}\sqrt{\frac{\alpha\beta k}{2j}} \right]d^{\dagger}d - \left[ (g_1+g_2)\frac{\beta k}{2j} - \omega\sqrt{\alpha} \right]\left( c^{\dagger} + c \right) + \left[ \frac{2(g_1+g_2)}{k}\sqrt{\frac{\alpha k}{2j}}(j - \beta) - \omega_0\sqrt{\beta} \right]\nonumber\\
    && \times\left( d^{\dagger} + d \right) + \frac{(g_1+g_2)}{4k^2}\sqrt{\frac{\alpha\beta k}{2j}}(2k + \beta)\left( d^{\dagger} + d \right)^2 + \frac{(g_1+g_2)\beta}{2k}\sqrt{\frac{k}{2j}}\left( c^{\dagger} + c \right)\left( d^{\dagger} + d \right) + \sqrt{\frac{k}{2j}}\Big[ g_1( c^{\dagger}d + c d^{\dagger} )\nonumber\\
&&  + g_2( c^{\dagger}d^{\dagger} + c d ) \Big] + \left[ \omega_0(\beta - j) + \omega \alpha - 2\frac{(g_1+g_2)}{k}\sqrt{\frac{\alpha\beta k}{2j}}\right]
  \end{eqnarray}
\end{widetext}
\begin{figure*}[t]
  \subfigure{\includegraphics[width=0.238\textwidth]{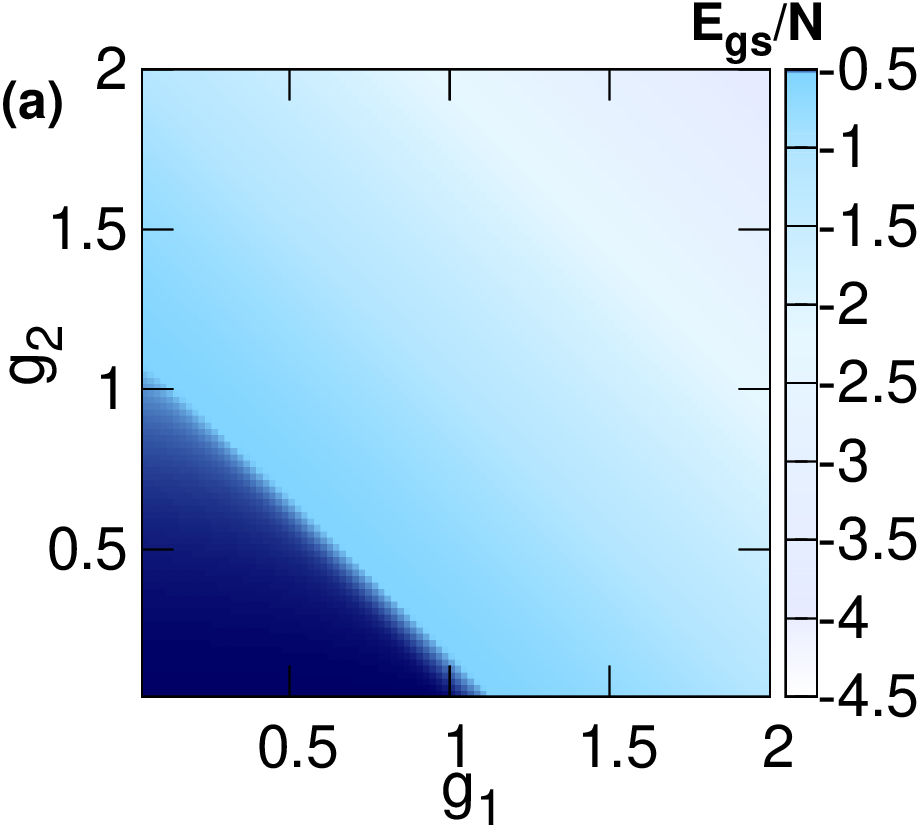}\label{fig:gs_energy_density}}
  \subfigure{\includegraphics[width=0.228\textwidth]{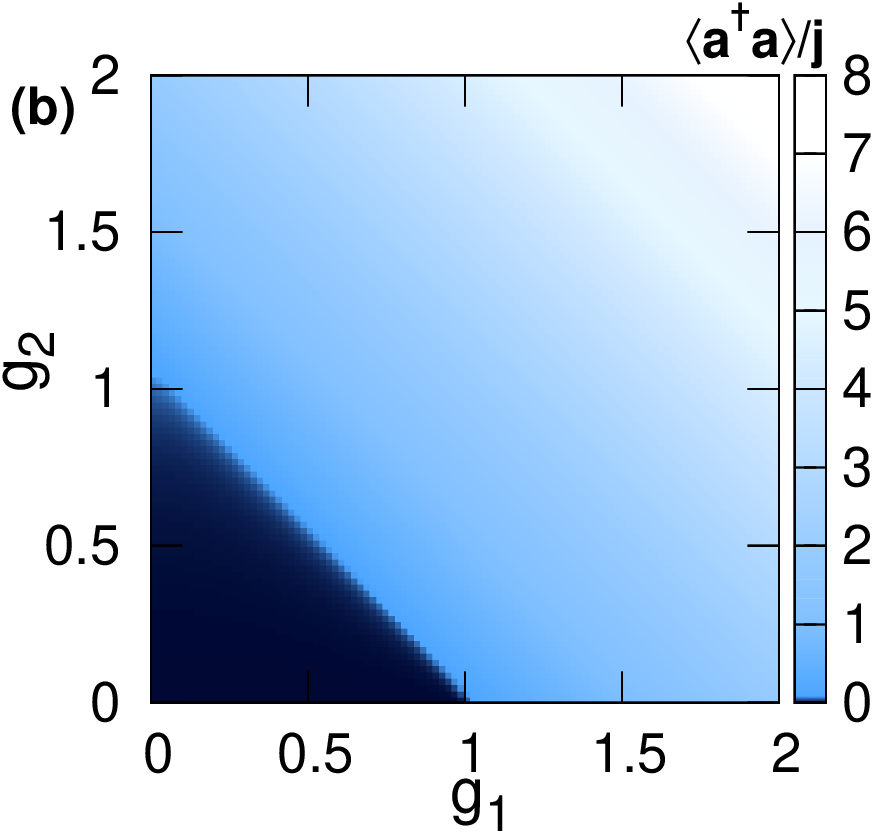}\label{fig:num_gs}}
  \subfigure{\includegraphics[width=0.248\textwidth]{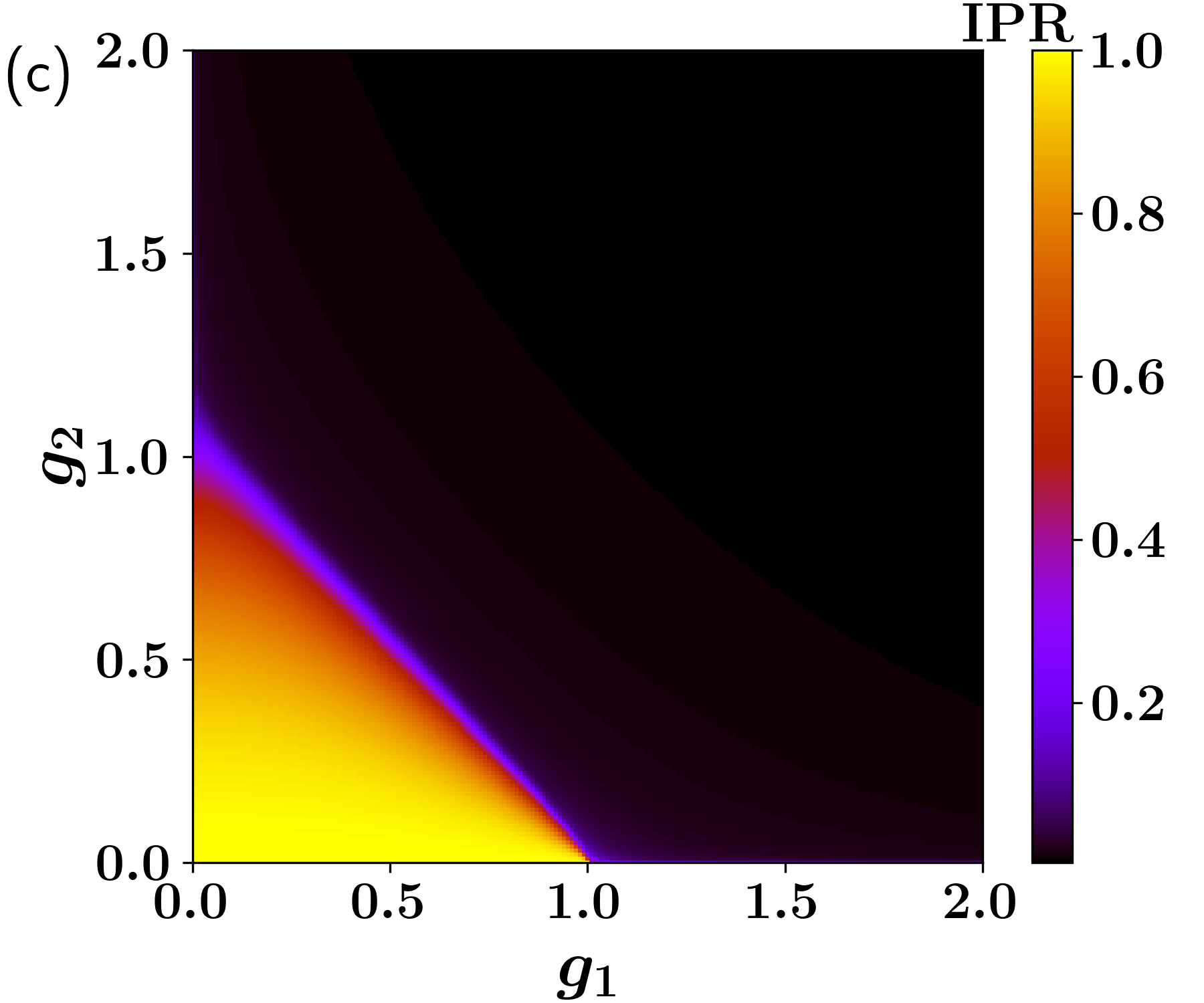}\label{fig:IPR_gs}}
  \subfigure{\includegraphics[width=0.227\textwidth]{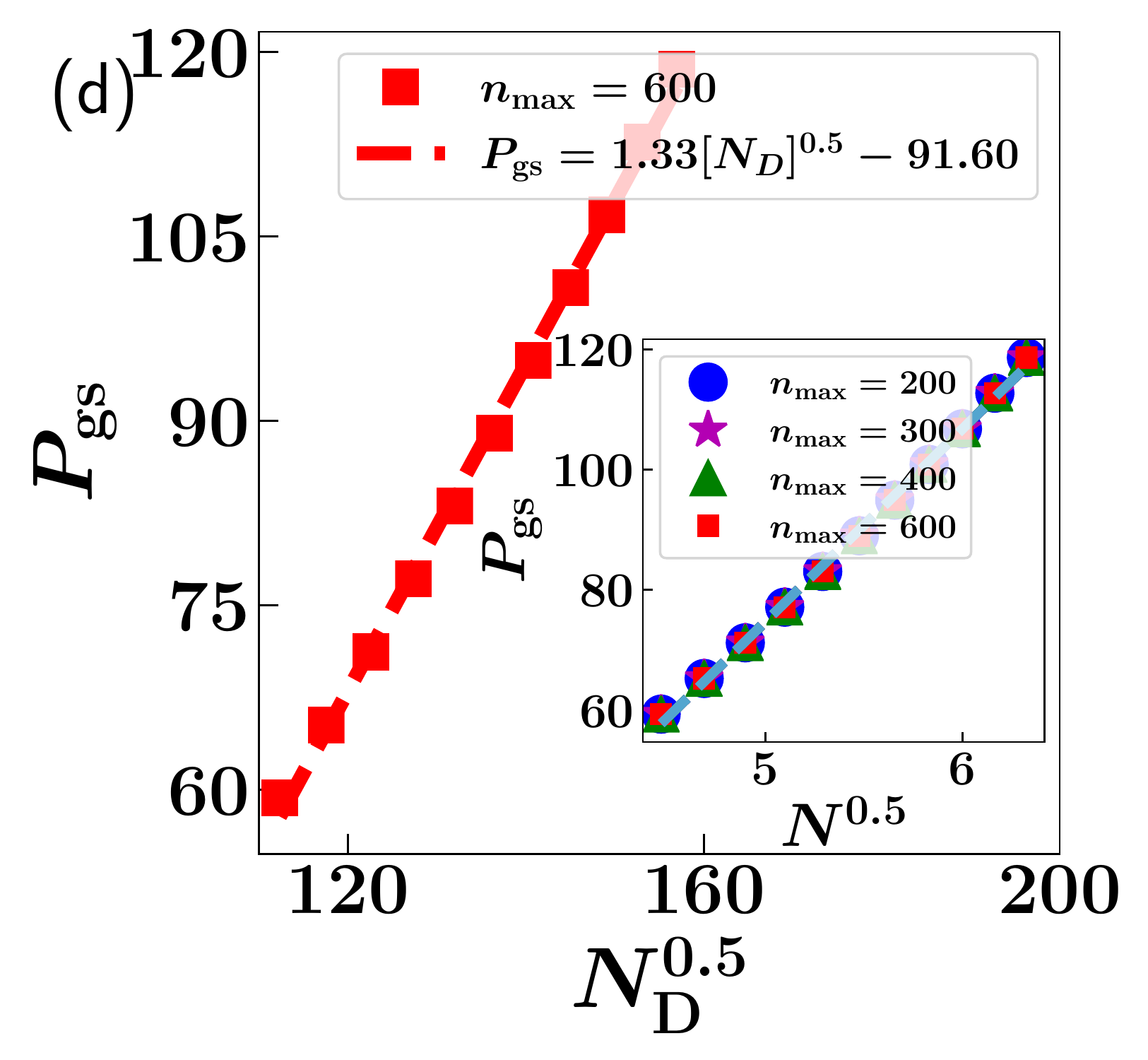}\label{fig:PR_gs}}
  \caption{ (a) Ground state energy of ADM shows NP to SP QPT. (b)
    Average number of boson (which is scaled by the pseudospin length
    $j$) is close to zero for NP ($g_1 + g_2 < 1$) and nonzero for SP
    ($g_1 + g_2 > 1$). (c) The ground state IPR is one in NP
    (localized) and close to zero in the SP (delocalized) (d) Scaling
    of ground state $PR$ with the full Hilbert space dimension $N_D$
    at some point $g_1=1.2$, $g_2=0.8$, in the SP for ADM for
    $n_{\text{max}}=600$ and changing the atom number: $N\in[20,
      40]$. $PR_{\text{gs}}$ scales as $\sqrt{N_D}$. Here $N_D =
    (n_{\text{max}} + 1)( N + 1)$ is increasing due to increase of
    atom number $N$. In the inset we consider four $n_{\text{max}}$
    values: $n_{\text{max}}=200$ (blue solid circles), $300$ (purple
    stars), $400$ (green triangles), $600$ (red squares) and changing
    the atom number: $N\in[20, 40]$. $PR_{\text{gs}}$ scales as
    $\sqrt{N}$. For each fixed $n_{\text{max}}$ dashed lines indicate
    the fitting with $\sqrt{N}$. The four sets of $P_{\text{gs}}$
    values are overlap. For (a)-(c) the parameters are: $\omega =
    \omega_0 = 1$, $N=40$. The bosonic cut-off is set to be $n_{\text{max}} =
    200$, and this is shown to be large enough. 
      Energy is calculated in units of $\omega_0$ and we fix
      $\omega_{0}=1$ throughout this paper.}
  \label{fig:qpt1}
\end{figure*}
where $k \equiv 2j - \beta$. This effective Hamiltonian in SP 
is also bilinear in the bosonic operators. The global symmetry 
$\Pi$ is broken at the phase transition and two new local 
symmetries appear corresponding to the operator $\Pi^{(2)} = 
\exp\Big( i\pi\Big[ c^{\dagger}c + d^{\dagger}d \Big] \Big)$~\cite{emary2003chaos}.

\section{QPT, ESQPT and ENET}\label{sec_3}

In this section we discuss three types of transitions in the
anisotropic Dicke model: QPT, ESQPT and ENET, in separate subsections.
To do the numerics we perform exact diagonalization of the system
Hamiltonian. Due to the bosonic mode, the Hilbert space dimension of
the ADM is infinite dimensional, however for numerics one has to
truncate the Hilbert space by cutting off the bosonic mode at some
finite $n_{\text{max}}$. We checked that our results remain
robust on increasing $n_{\text{max}}$ (see the Appendix).

\subsection{Quantum phase transition (QPT)}
In Fig.~\ref{fig:gs_energy_density} we show the ground state energy density
(energy scaled by the atom number, $E_{\text{gs}}/N$) of the system
Hamiltonian as a function of $g_1$ and $g_2$. While in the normal
phase, the energy is almost constant close to $=-0.5$ (which is very
similar to the value seen for the isotropic Dicke model), the
super-radiant phase has a broad energy spectrum with the density
ranging from $-4.5\leq \frac{E_{gs}}{N}\leq -0.5$.  This clearly
distinguishes the normal and super-radiant phases. Furthermore, the
mean photon number given by the operator $\langle a^{\dagger}
a\rangle/j$~\cite{emary2003chaos} is almost zero in the normal phase,
while in the super-radiant phase it has a non-zero value with a
continuous change across the transition line $g_1 + g_2 = 1$
(Fig.~\ref{fig:num_gs}), and indicates a second order phase transition.

Finally, we study the ground state properties by looking at the degree
of localization using the (inverse) participation ratio. The
participation ratio~\cite{edwards1972numerical, roy2020interplay} (PR)
of an eigenstate $\vert\psi\rangle =\sum_j^{N_D}\psi_j\vert j\rangle$
(where $N_D$ is the Hilbert space dimension) is defined as:
\begin{equation}\label{eq:PR}
  P = \frac{1}{\sum_{j=1}^{N_D}\vert\psi_j\vert^{4}}.
\end{equation}
The inverse of the participation ratio, called the inverse
participation ratio ($\text{IPR} = 1/\text{PR}$) is often a useful
measure in its own right. From the inset of Fig.~\ref{fig:IPR_gs}, it
is clear that, in the normal phase, IPR is close to unity suggesting
that the ground state is localized. On the other hand, a careful study
of the scaling of PR in the super-radiant phase with the Hilbert-space
dimension $N_D$ reveals interesting features. The PR scales as
PR~$\sim \sqrt{N_D}$ and suggests that the ground state in the
super-radiant phase exhibits multi-fractal behavior (see
Fig.~\ref{fig:PR_gs}).

  In Fig.~\ref{fig:PR_gs} we fix the parameters to be
  $g_1 = 1.2, g_2 = 0.8$ to show a representative example in the
  SP. While the bosonic cut-off is fixed at $n_{\text{max}} = 600$ we
  vary the atom number $N$ and hence the total Hilbert space dimension
  $N_D$ of the truncated model also varies. The change in $N_D$ is
  entirely due to the change in atom number $N$, since
  $n_{\text{max}}$ remains fixed.  In this figure the red squares are
  the data for $P_{\text{gs}}$ whereas the red dashed line denotes the
  fit to the functional form $N_D^{0.5}$. In the inset of
  Fig.~\ref{fig:PR_gs} we choose four different $n_{\text{max}}$
  values: $n_{\text{max}} = 200$ (blue solid circles), $300$ (purple
  stars), $400$ (green triangles), $600$ (red squares) and for each
  fixed $n_{\text{max}}$ we show data as the atom number $N$ (and
  hence the total Hilbert space dimension $N_D$) varies exactly like
  in the main figure. The four sets of data corresponding to different
  $n_{\text{max}}$ exactly overlap with each other.  The dashed lines
  denote the data fitting of $P_{\text{gs}}$ with $N^{0.5}$.  Hence we
  can conclude that for a fixed atom number $N$, $P_{\text{gs}}$ is
  independent of $n_{\text{max}}$, and only depends on $N$.

\subsection{Excited state quantum phase transition (ESQPT)}

\begin{figure*}[t]
  \subfigure{\includegraphics[width=0.205\textwidth]{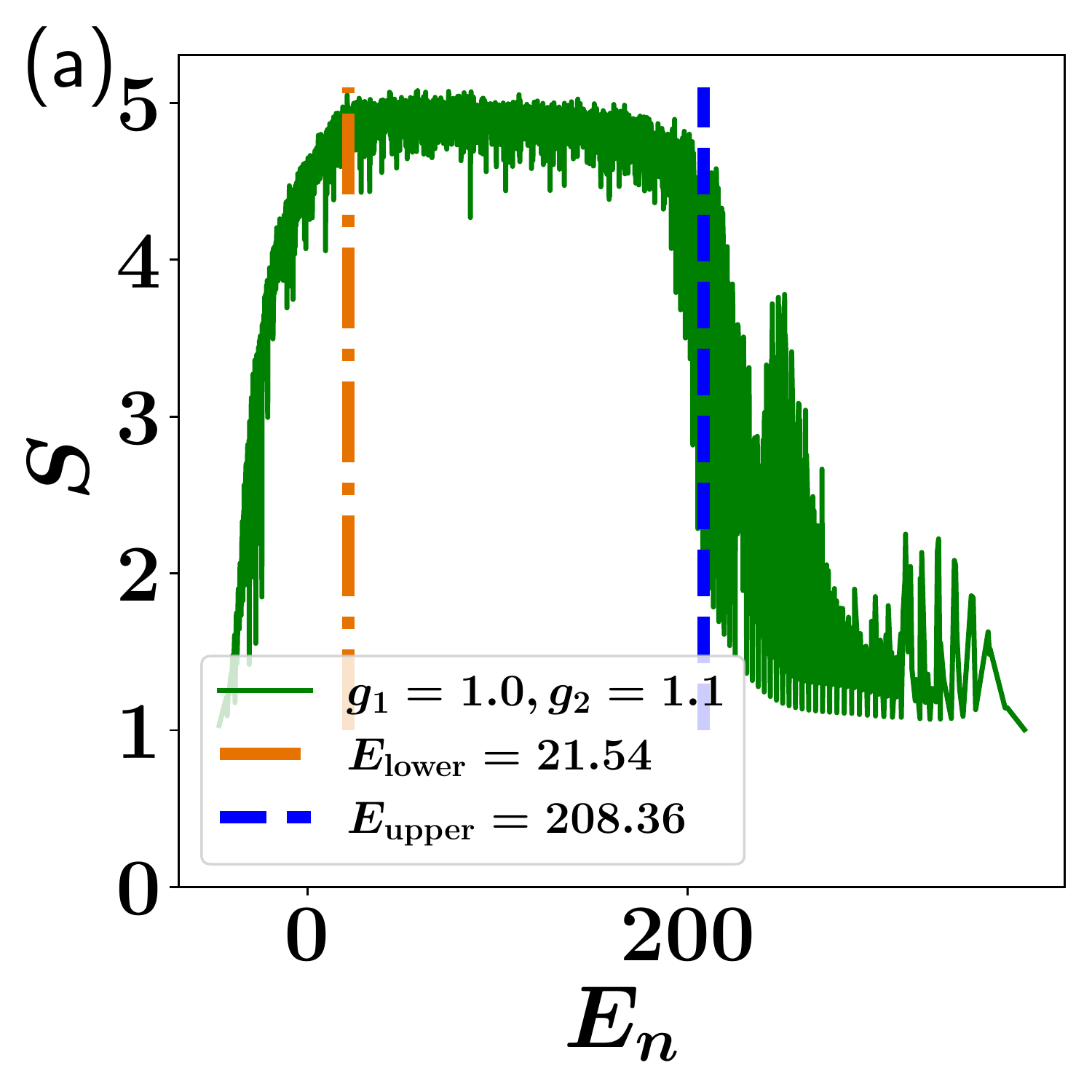}\label{fig:VNEE}}
  \subfigure{\includegraphics[width=0.235\textwidth]{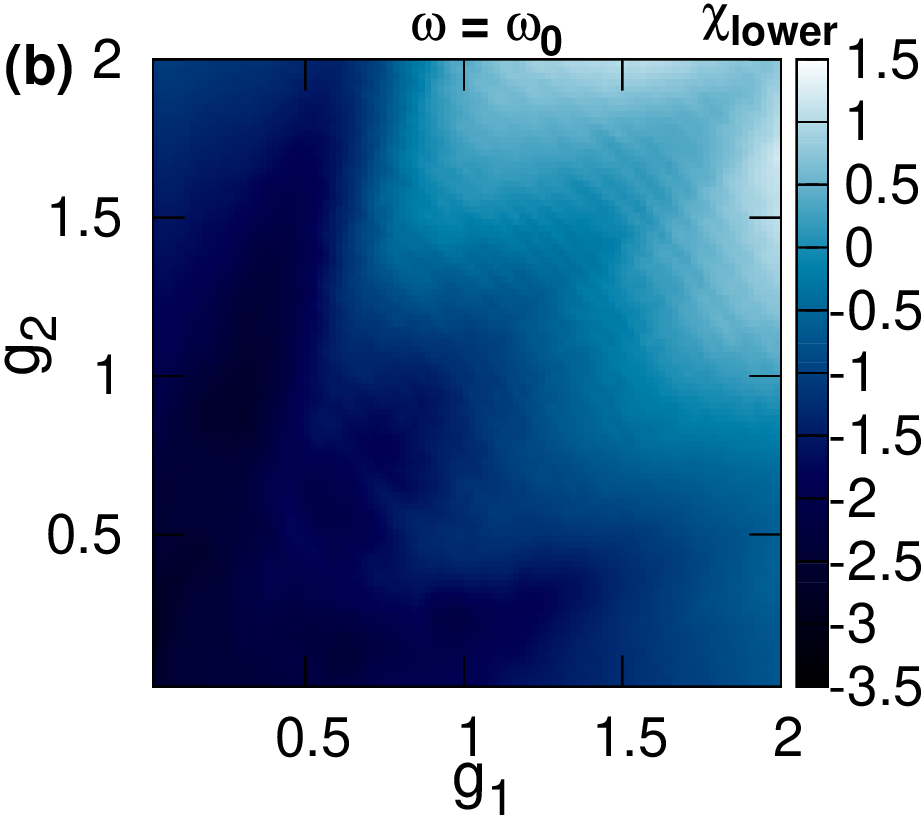}\label{fig:lower_energy}}
  \subfigure{\includegraphics[width=0.245\textwidth]{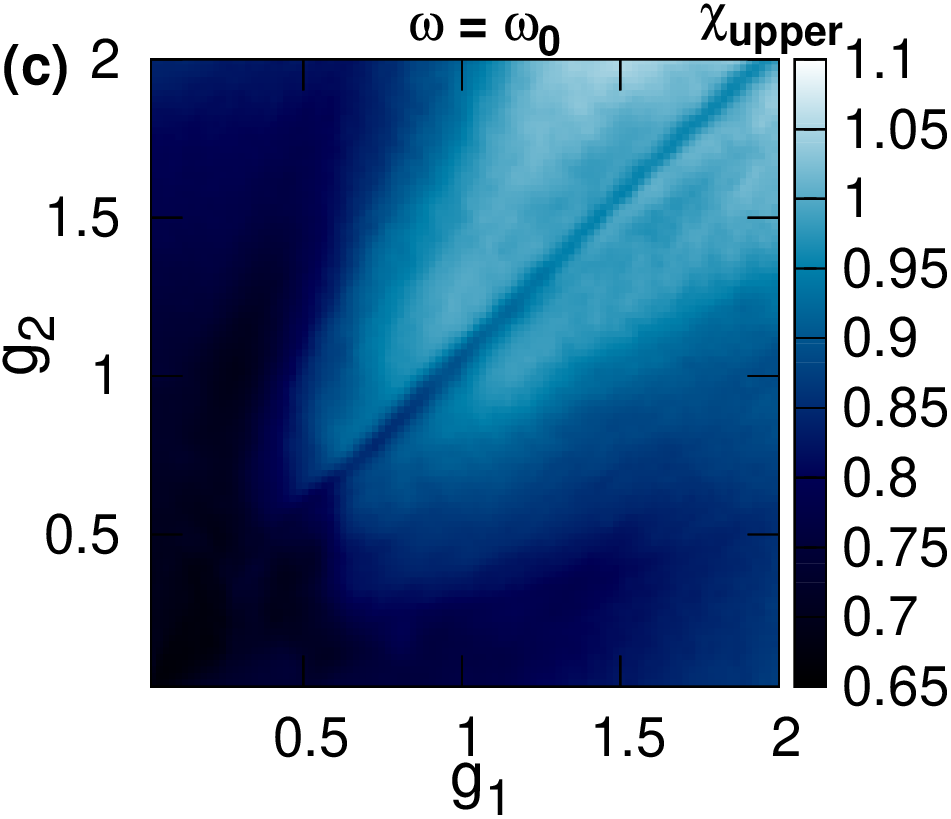}\label{fig:upper_energy}}
  \subfigure{\includegraphics[width=0.24\textwidth]{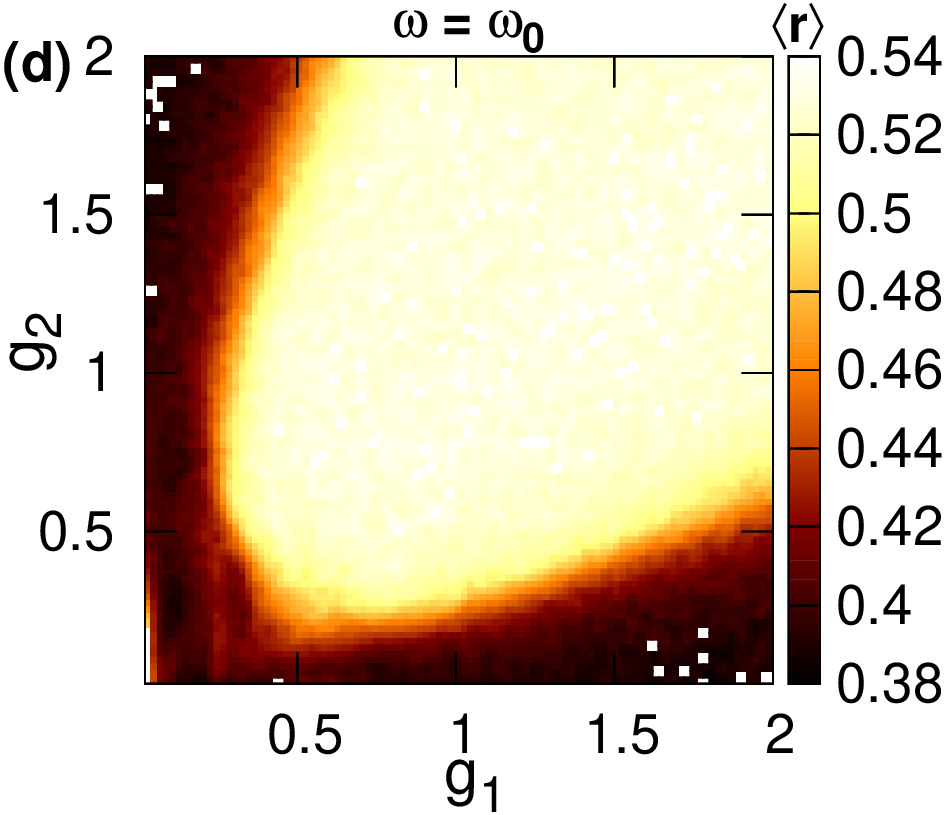}\label{fig:r_avg}}
  
  \subfigure{\includegraphics[width=0.21\textwidth]{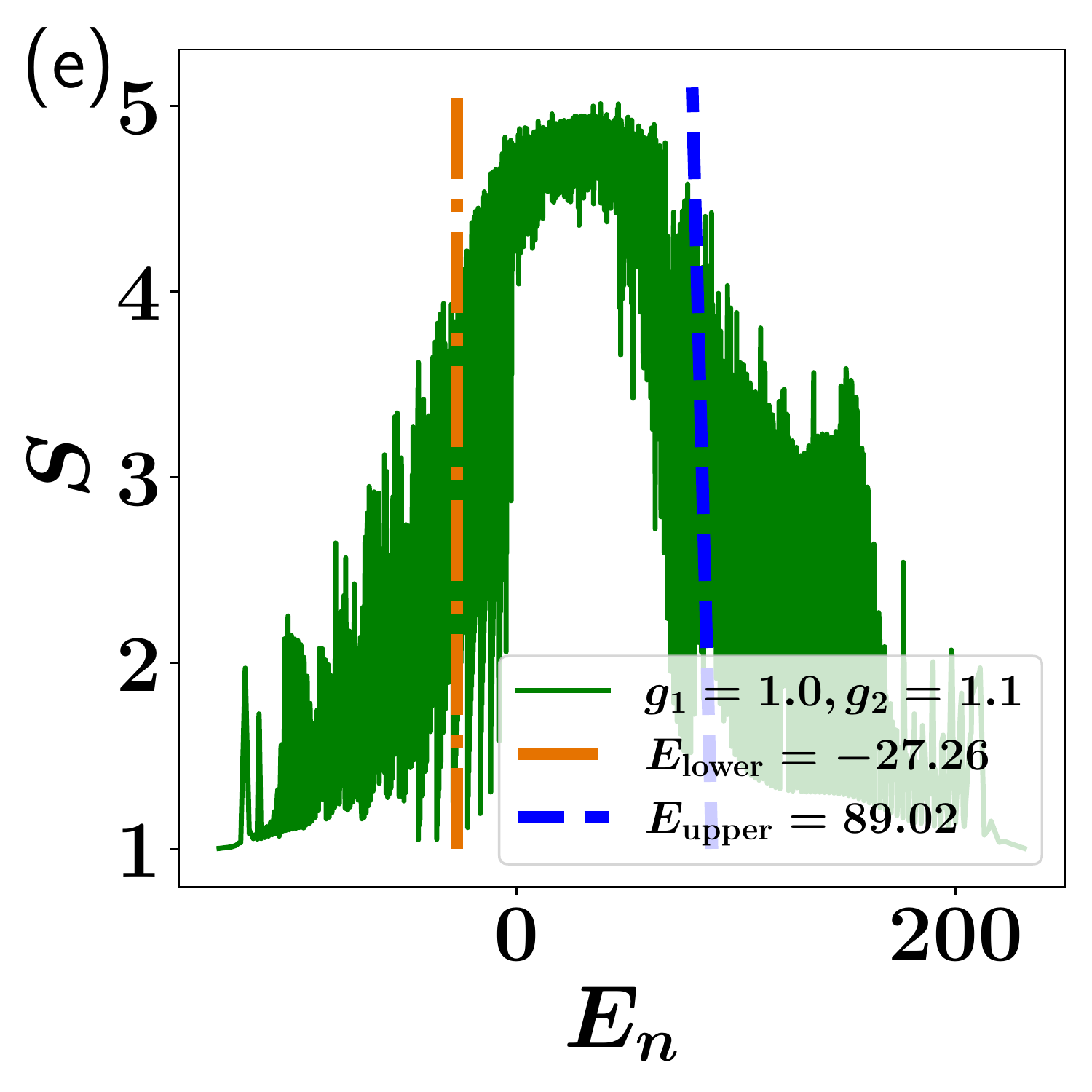}\label{fig:VNEE_off1}}
  \subfigure{\includegraphics[width=0.235\textwidth]{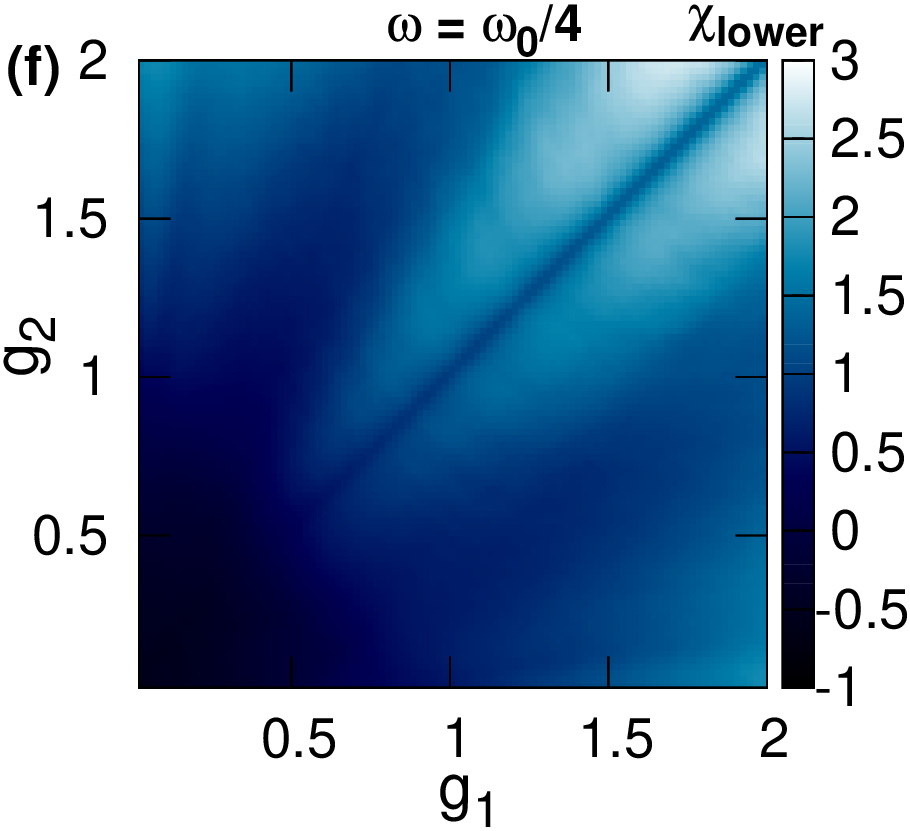}\label{fig:lower_energy_off1}}
  \subfigure{\includegraphics[width=0.235\textwidth]{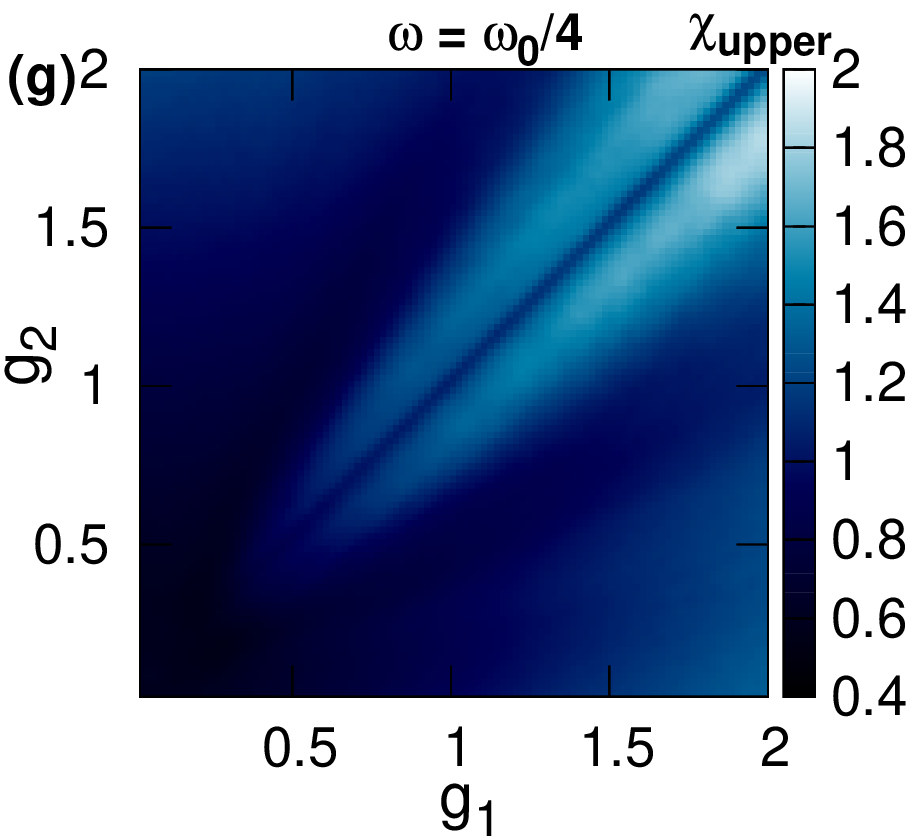}\label{fig:upper_energy_off1}}
  \subfigure{\includegraphics[width=0.25\textwidth]{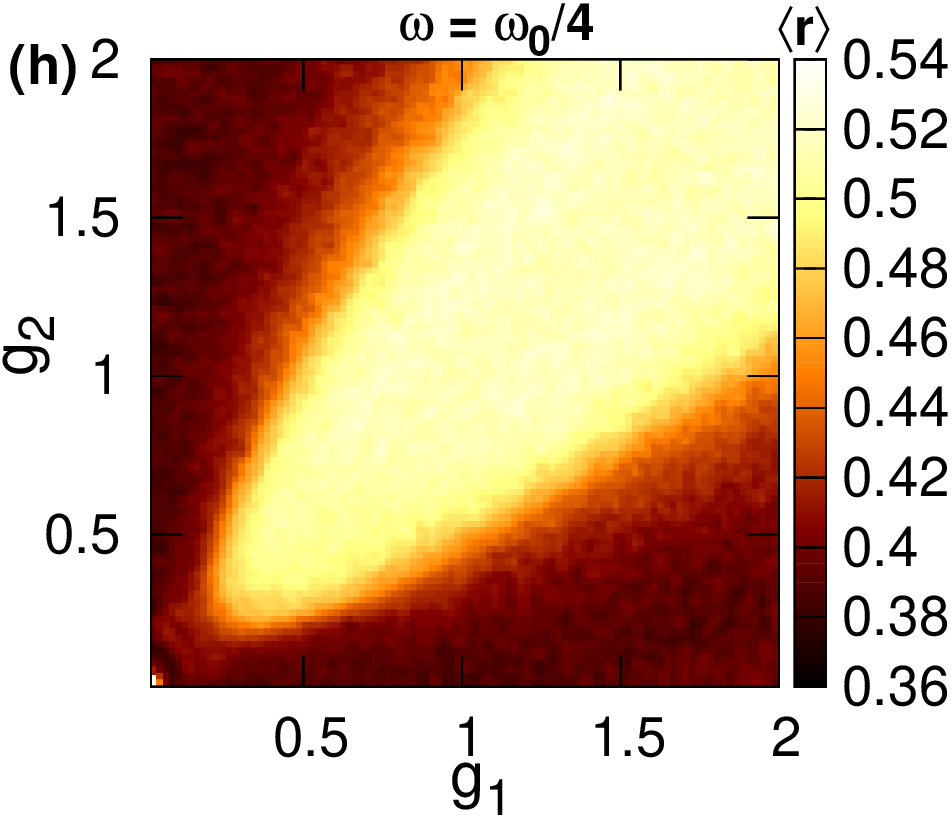}\label{fig:r_avg_off1}}
  
  \subfigure{\includegraphics[width=0.205\textwidth]{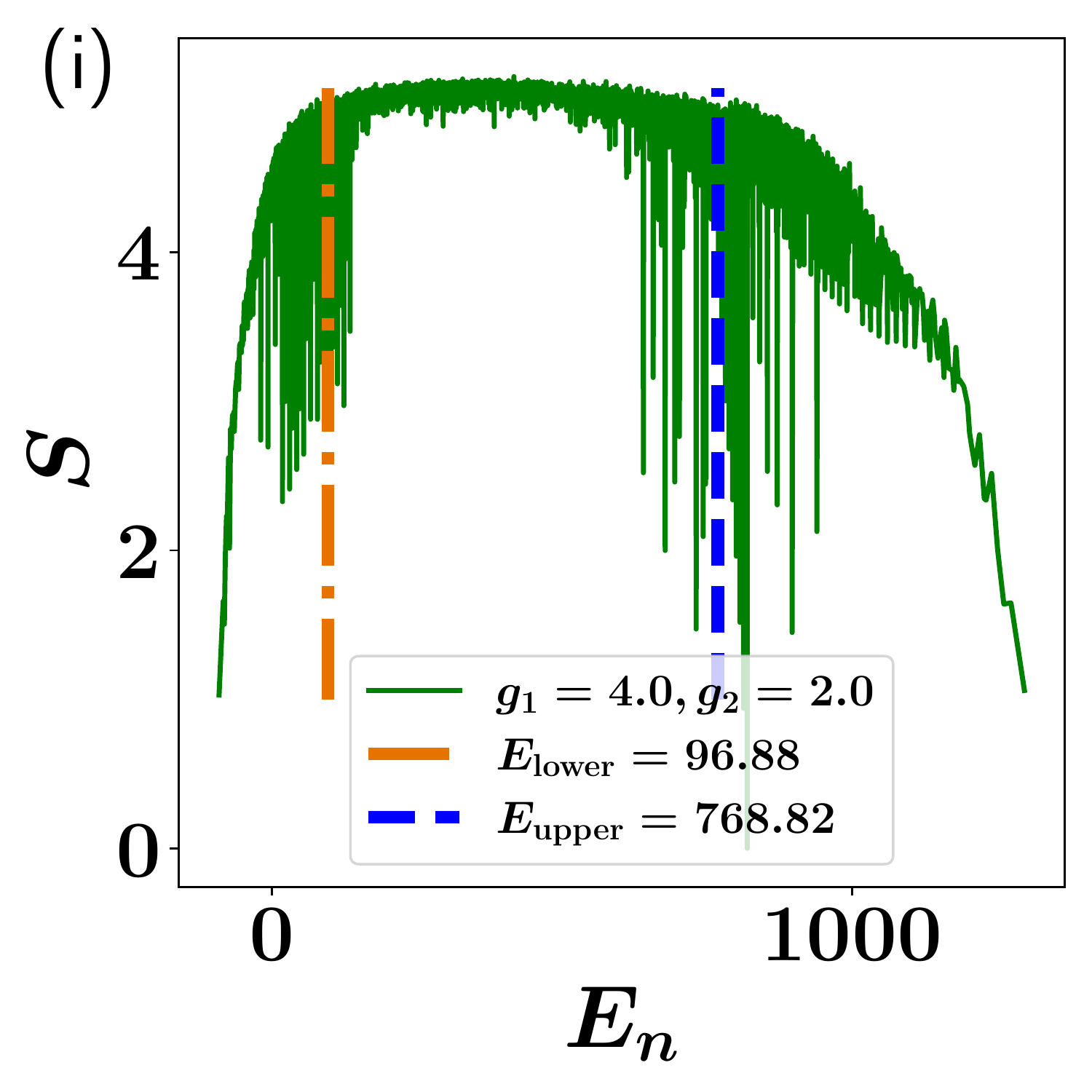}\label{fig:VNEE_off2}}
  \subfigure{\includegraphics[width=0.23\textwidth]{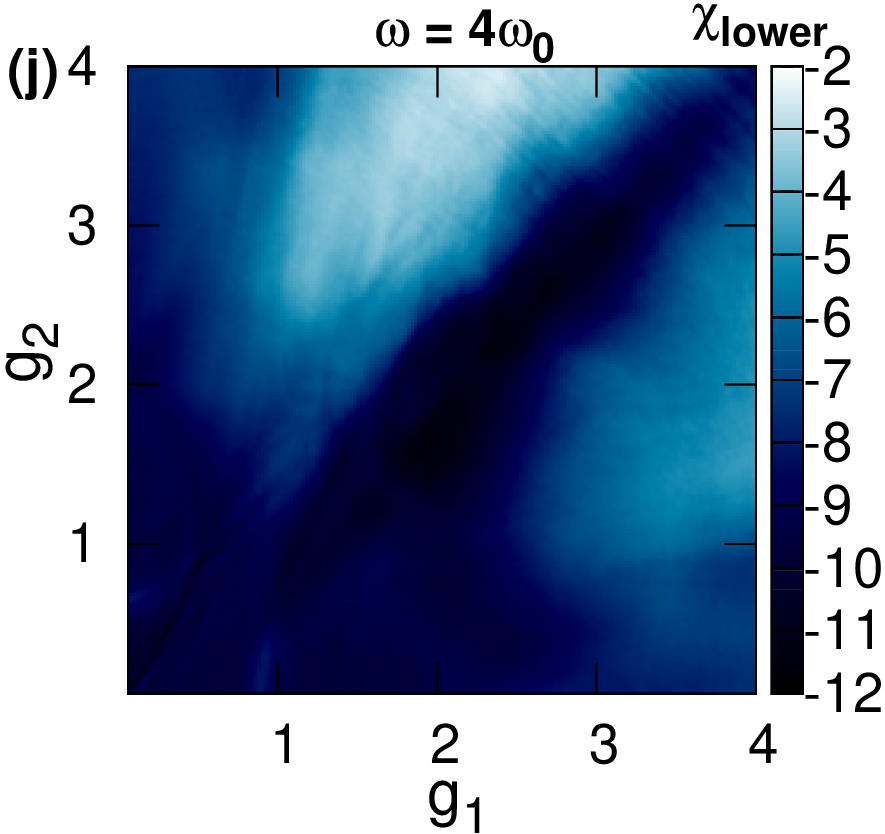}\label{fig:lower_energy_off2}}
  \subfigure{\includegraphics[width=0.24\textwidth]{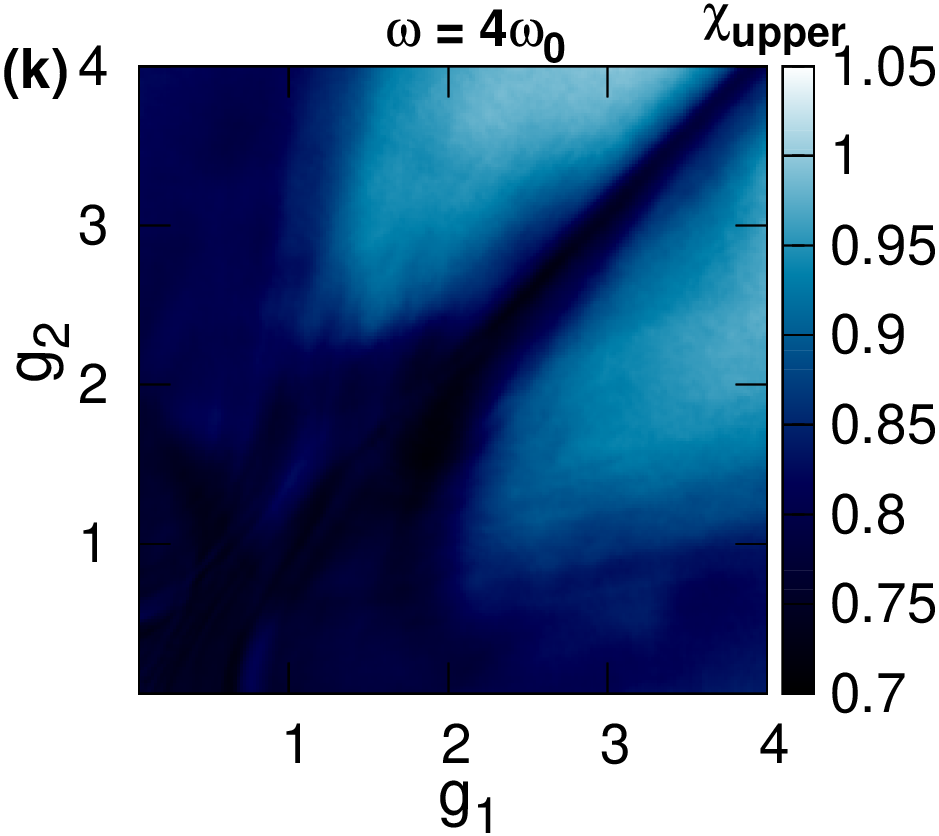}\label{fig:upper_energy_off2}}
  \subfigure{\includegraphics[width=0.24\textwidth]{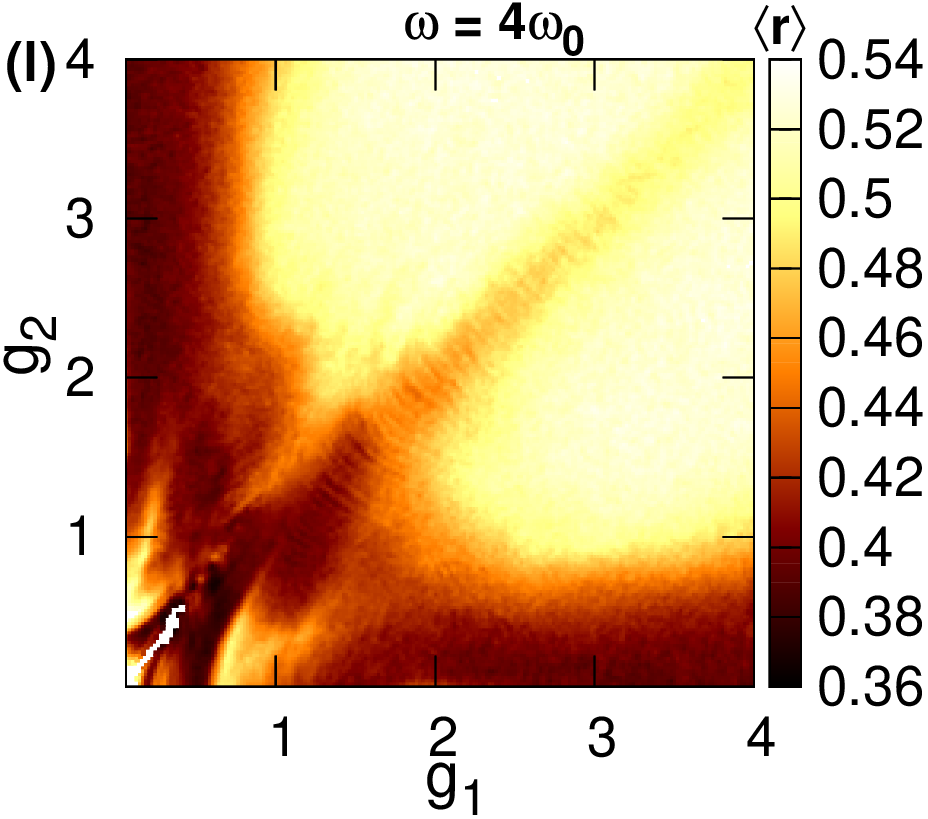}\label{fig:r_avg_off2}}
  \caption{(a), (e), and (i) Von-Neumann entanglement entropy as a
    function of the eigenstate energies of the ADM Hamiltonian at the
    point $g_1=1.0,\ g_2=1.1$ [for (a) and (e)] and at the point $g_1
    = 4,\ g_2 = 2$ [for (i)]. In the figure the orange dash-dotted
    line denotes $E_{\text{lower}}$ [given in Eqn.~\ref{eqn:E_lower}]
    and the blue dashed line is for $E_{\text{upper}}$ (given in
    Eqn.~\ref{eqn:E_upper}). (b), (f), and (j) $\chi_{\text{lower}}$
    [given in Eqn.~\ref{eqn:chi_lower}] as a function of $g_1$ and
    $g_2$, (c), (g), and (k) $\chi_{\text{upper}}$ [given in
    Eqn.~\ref{eqn:chi_upper}] as a function of $g_1$ and $g_2$. (b)
    and (c) shows similar phase transition from non-ergodic to ergodic
    phase, which suggest that ESQPT is related to ENET. (d), (h), and (l)
    Consecutive level spacing ratio, $\langle r\rangle$ of the
    energies lying between $E_{\text{normal}}^{0}$,
    $E_{g_1,g_2=0}^{\text{max}}$. For all the above figures atom
    number, $N=40$, and we take the bosonic cut-off: $n_{\text{max}} =
    200$. Top panel is for resonant case $\omega = \omega_0 = 1$,
    middle and bottom panels are for two off-resonant cases: $\omega =
    \frac{\omega_0}{4}$ and $\omega = 4\omega_0$ respectively. }
  \label{fig:esqpt}
\end{figure*}   

We now look at the properties of the excited states and find that
similar to the Dicke model, the anisotropic Dicke model also exhibits
an excited state quantum phase transition in the super-radiant phase
(for $g_1+g_2>1$). While the literature on ESQPT has mainly considered
eigenvalue properties~\cite{perez2011excited, lewis2019unifying,
  perez2017thermal, brandes2013excited, bastarrachea2014comparative,
  stransky2016classification, stransky2017excited}, we showned in
our recent paper~\cite{das2022revisiting} the profitability of
studying eigenvector properties such as the von Neumann entanglement
entropy~\cite{ lambert2004entanglement}, the average bosonic
number~\cite{ emary2003chaos}, concurrence~\cite{hill1997entanglement,
  wootters1998entanglement} and participation ratio~\cite{
  edwards1972numerical, roy2020interplay}. We also aruged that
there is not only a lower cut-off energy but also an upper cut-off and
energies between these two cut-offs behave differently from the upper
and lower bands. In the Dicke model, the lower cut-off energy~\cite{
  perez2011excited, lewis2019unifying} is the ground state energy at
the critical coupling strength $g_c$ and the upper cut-off
energy~\cite{das2022revisiting} is the maximum energy at $g=0$ (for
finite $n_{\text{max}}$). On the other hand, in the ADM, we find that
the lower cut-off energy is around the ground state energy of the
system on the critical line of quantum phase transitions ($g_1+g_2=1$)
while the upper cut-off energy is the maximum energy at $g_1=g_2 =0$
(for finite $n_{\text{max}}$). Energies below (above) the lower
(upper) cut-off form the lower (upper) energy band and the energies in
between the two cut-offs form the central band.

The entanglement entropy between the spins and the bosons is simply
the von Neumann entropy of the reduced density matrix of the spins:
\begin{equation}
  S_{\text{spins}} = -\text{Tr}[\rho_{\text{spins}} \ln(\rho_{\text{spins}})]
\label{eqn:VNEE}
\end{equation} 
where $\rho_{\text{spins}}$ is the reduced density matrix of the spins
obtained by tracing over the bosonic degrees of freedom.  In
Fig.~\ref{fig:VNEE} we show the von Neumann entanglement entropy
(between spins and bosons) as a function of the energy eigenvalues at
some point $g_1=1.0$ and $g_2=1.1$. We observe from this
representative plot that there are three characteristic parts: (i) an
increasing part upto some energy, followed by (ii) a plateau region,
and then (iii) a region where VNEE decreases. To quantify the
beginning and the end of the plateau region, we define two
characteristic energies $E_{\text{lower}}$ and $E_{\text{upper}}$ as:
\begin{equation}
  E_{\text{lower}} = \left[ \frac{\sum_{n=0}^{\frac{N_D}{2}}E_n|\Delta S_n|}{\sum_{n=0}^{\frac{N_D}{2}}|\Delta S_n|} \right]
  \label{eqn:E_lower}
\end{equation} 
and
\begin{equation}
  E_{\text{upper}} = \left[ \frac{\sum_{n=\frac{N_D}{2}}^{N_D}E_n|\Delta S_n|}{\sum_{n=\frac{N_D}{2}}^{N_D}|\Delta S_n|} \right]
  \label{eqn:E_upper}
\end{equation}
where $\Delta S_n = S_{n+1}-S_n$ is the VNEE difference between that
of the $(n+1)^{\text{th}}$ eigenstate and the $n^{\text{th}}$
eigenstate, and $E_n$ is the $n^{\text{th}}$ energy. The above
energies are obtained by using the jumps in the VNEE as weights. The
change in VNEE is taken as weights for the different energies, and we
would thus expect these quantities to signal the two ends of the
plateau region. In Fig.~\ref{fig:VNEE}, these two energies are marked
by two vertical lines: the orange dash-dotted line denotes
$E_{\text{lower}}$ whereas the blue dashed line denotes
$E_{\text{upper}}$ and they more or less match with the two end
energies of the plateau region. Now we define two more new quantities,
$\chi_{\text{lower}}$ and $\chi_{\text{upper}}$:
\begin{equation}
  \chi_{\text{lower}} = \frac{ E_{\text{lower}}}{E_{\text{normal}}^{0}}
  \label{eqn:chi_lower}
\end{equation}
\begin{equation}
  \chi_{\text{upper}} = \frac{ E_{\text{upper}}}{E_{g_1,g_2=0}^{\text{max}}}
    \label{eqn:chi_upper}
\end{equation}
where $E_{\text{normal}}^{0}$ is the minimum energy in the normal
phase ($g_1+g_2<1$) and $E_{g_1,g_2=0}^{\text{max}}$ is the maximum
energy at $g_1=g_2=0$ (for finite $n_{\text{max}}$). We plotted
these two quantities as a function of $g_1$ and $g_2$ in
Fig.~\ref{fig:lower_energy} and
\ref{fig:upper_energy}. Remarkably this reveals a clear visual
correlation between the ESQPT and what is called the ergodic to
non-ergodic phase transition (ENET)~\cite{buijsman2017nonergodicity},
which we describe in greater detail in the following
subsection. Moreover we notice that, along the diagonal line [it is
more clear in Fig.~\ref{fig:upper_energy}] there is a relatively dark
line which indicates that the symmetric Dicke model is special. A
quantitative way of identifying the central band of energies
corresponding to the plateau region in Fig.~\ref{fig:VNEE} is to
consider the energies that lie between $\chi_{\text{lower}}=1$ and
$\chi_{\text{upper}}=1$, i.e $E_{\text{lower}} =
E_{\text{normal}}^{0}$ and $E_{\text{upper}} =
E_{g_1,g_2=0}^{\text{max}}$. We perform a study of the average level
spacing ratio~\cite{atas2013distribution} $\langle r\rangle$ for the
energies lying between $E_{\text{lower}}$ and $E_{\text{upper}}$. Let
${s_n}$ denote the level spacing between two consecutive energies
$E_{n+1}$ and $E_n$, then the $\langle r \rangle$ is defined as the
average over $n$ of the ratio of consecutive level spacings:
\begin{equation}
  r_{n} = \frac{\text{min}(s_{n-1},\hspace{1mm}s_{n})}{\text{max}(s_{n-1},\hspace{1mm} s_n)}.
\end{equation} 
For an ergodic system, the value of $\langle r\rangle = 0.53$ and the probability 
distribution of the consecutive gaps shows Wigner-Dyson statistics~
\cite{kamenev1999wigner} 
while for a non-ergodic system  it is $\langle r\rangle = 0.386$ and the probability 
distribution becomes Poissonian.

    \begin{figure*}[htbp]
      \subfigure{\includegraphics[width=0.245\textwidth]{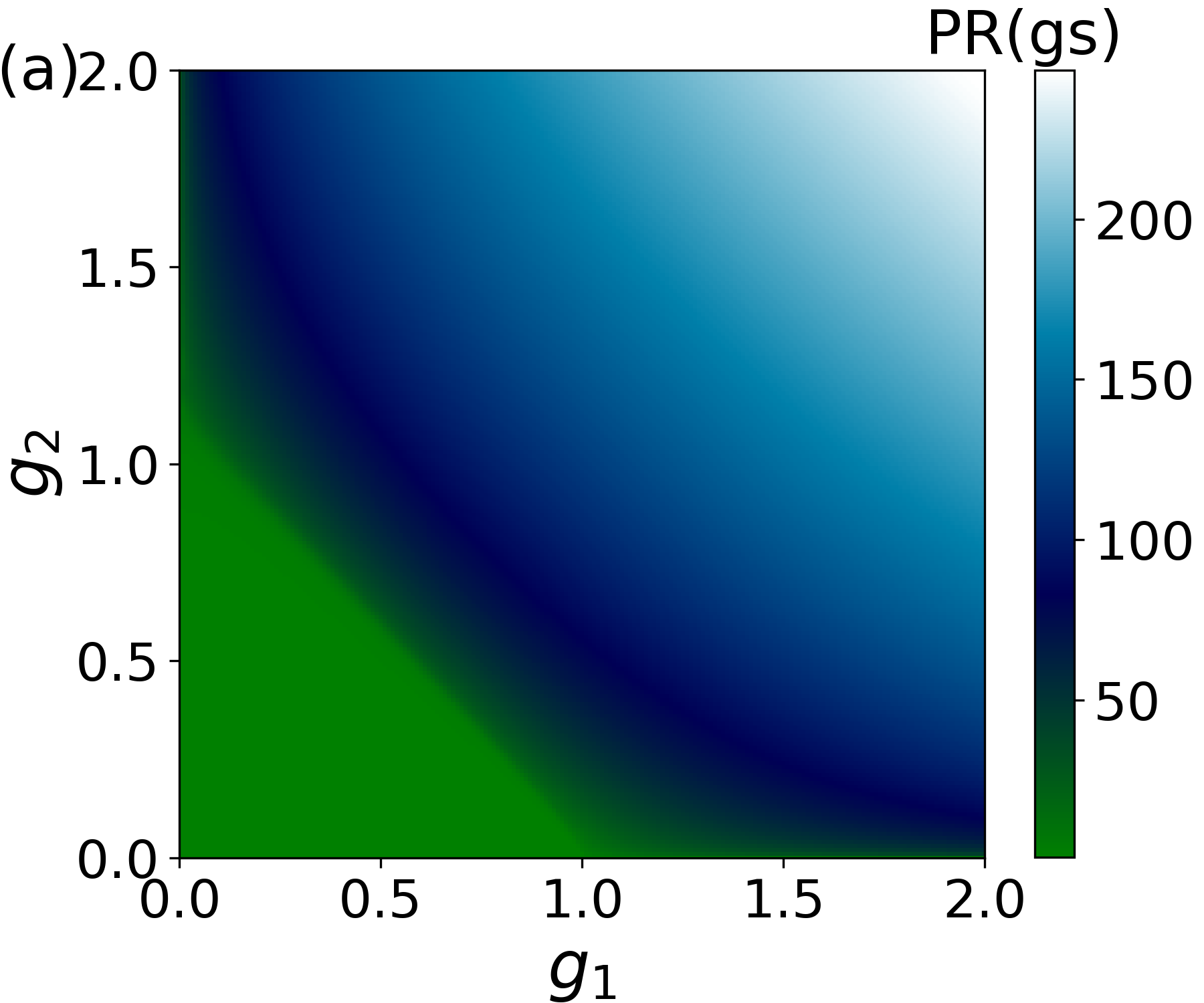}\label{fig:PR1}}
      \subfigure{\includegraphics[width=0.245\textwidth]{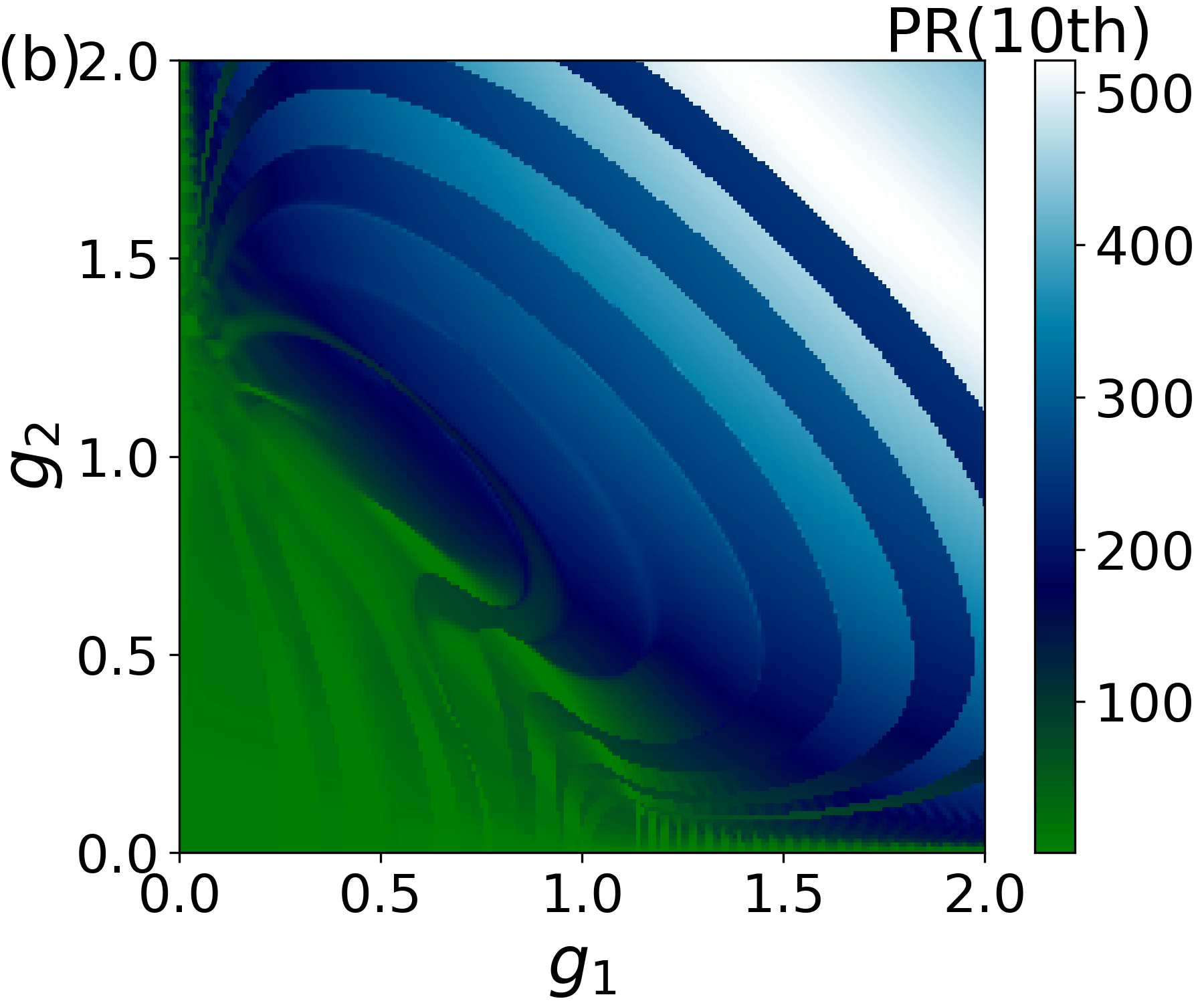}\label{fig:PR2}}
      \subfigure{\includegraphics[width=0.245\textwidth]{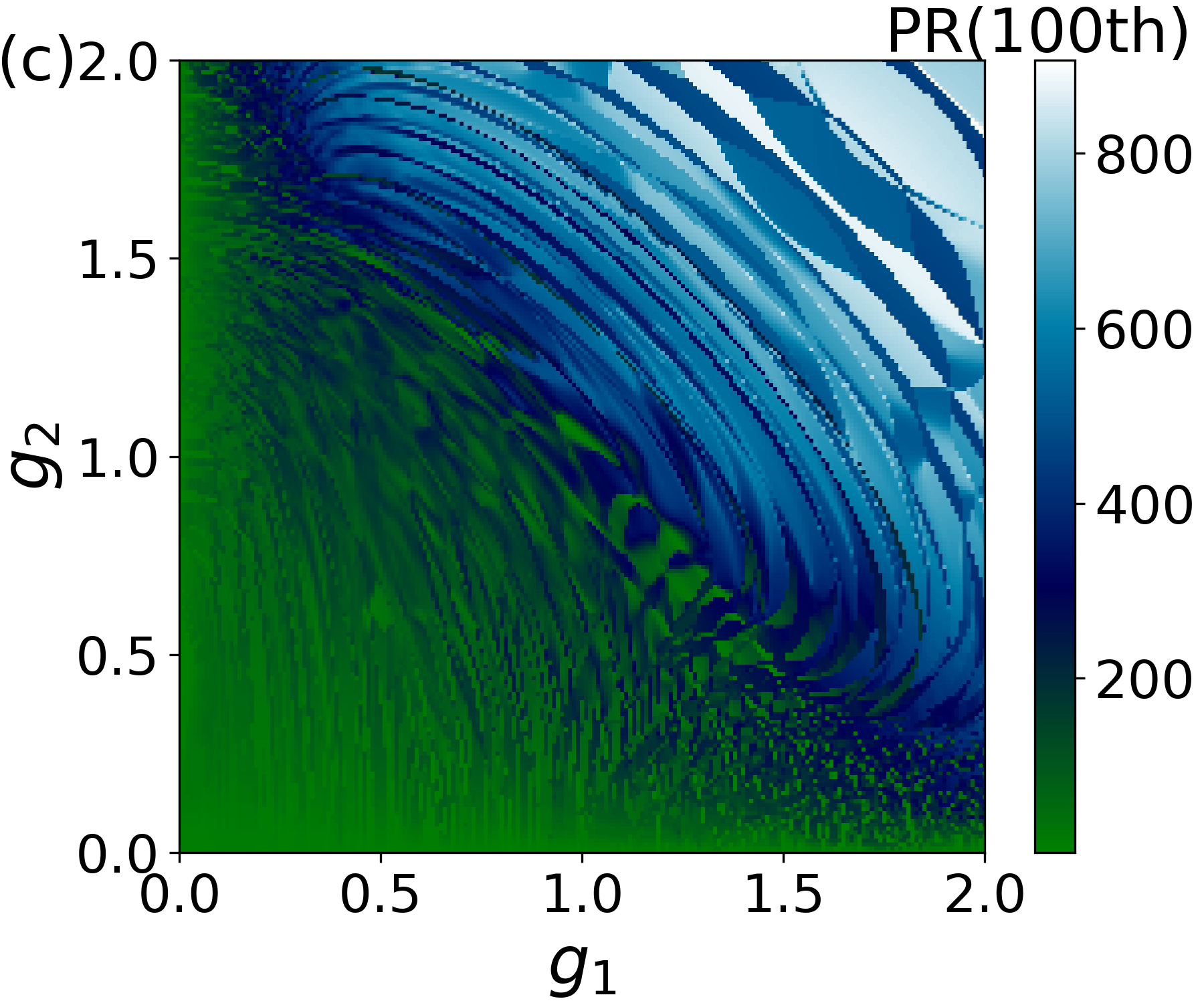}\label{fig:PR3}}
      \subfigure{\includegraphics[width=0.245\textwidth]{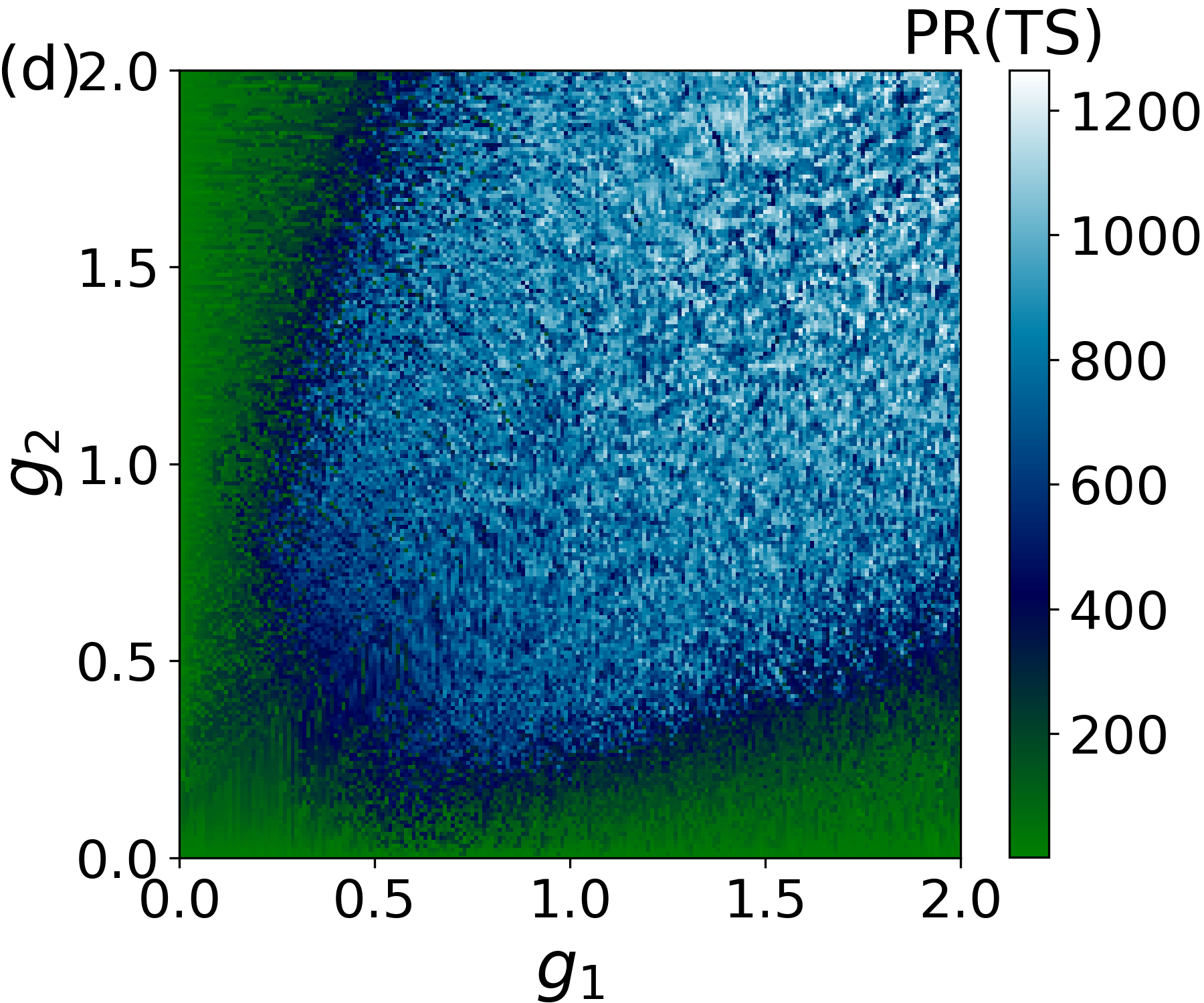}\label{fig:PR4}}
      \subfigure{\includegraphics[width=0.245\textwidth]{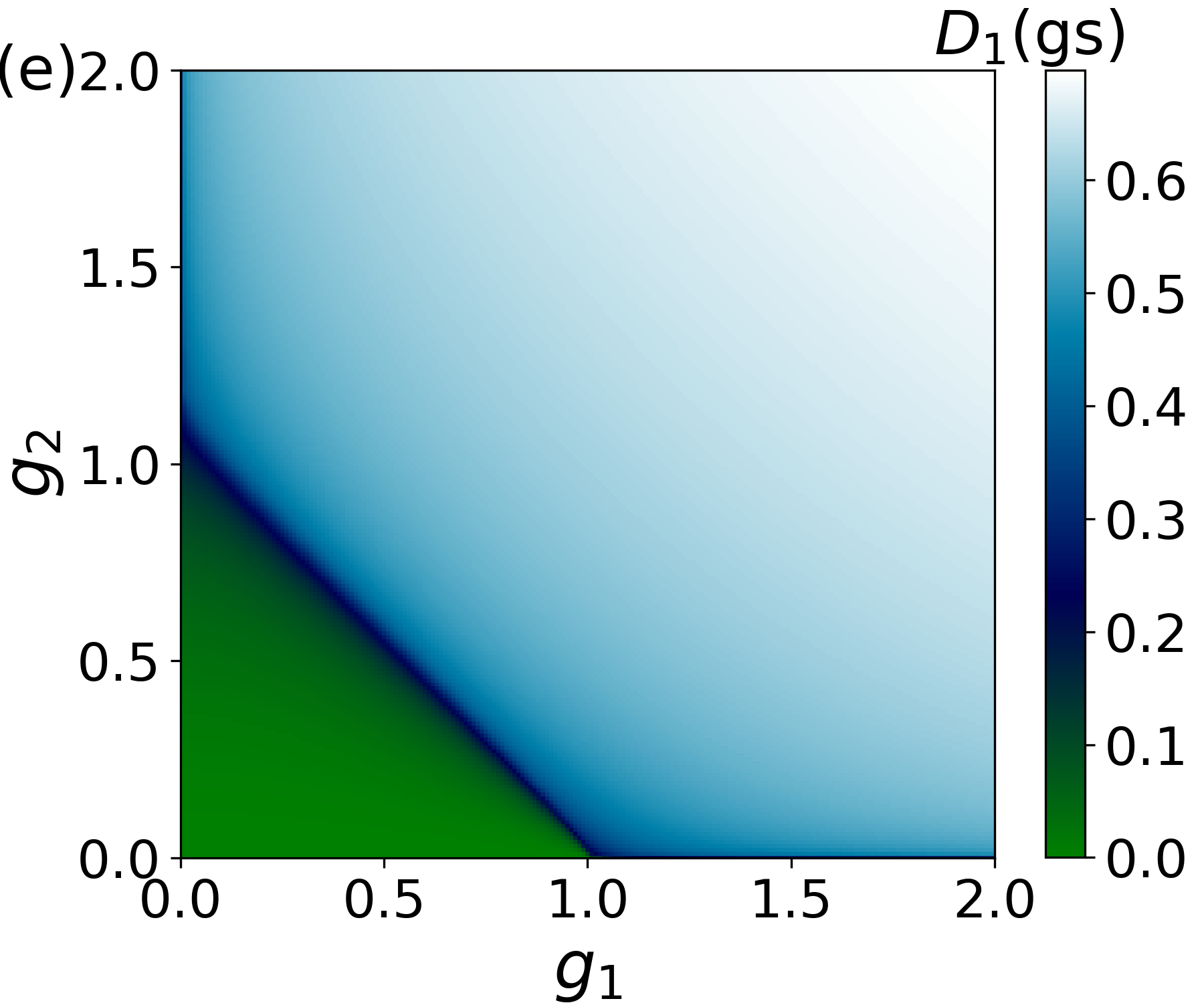}\label{fig:D1_GS}}
      \subfigure{\includegraphics[width=0.245\textwidth]{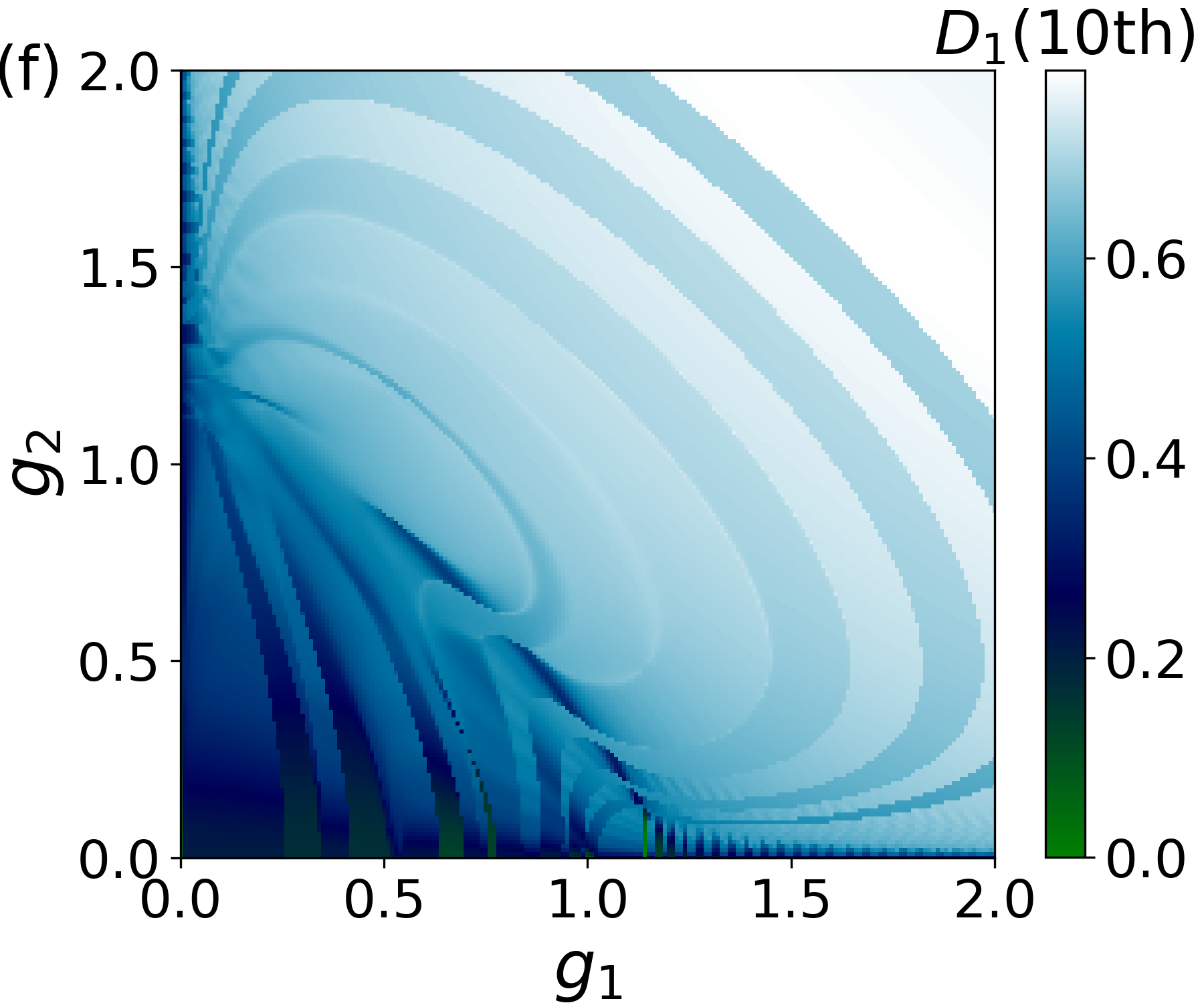}\label{fig:D1_10th}}
      \subfigure{\includegraphics[width=0.245\textwidth]{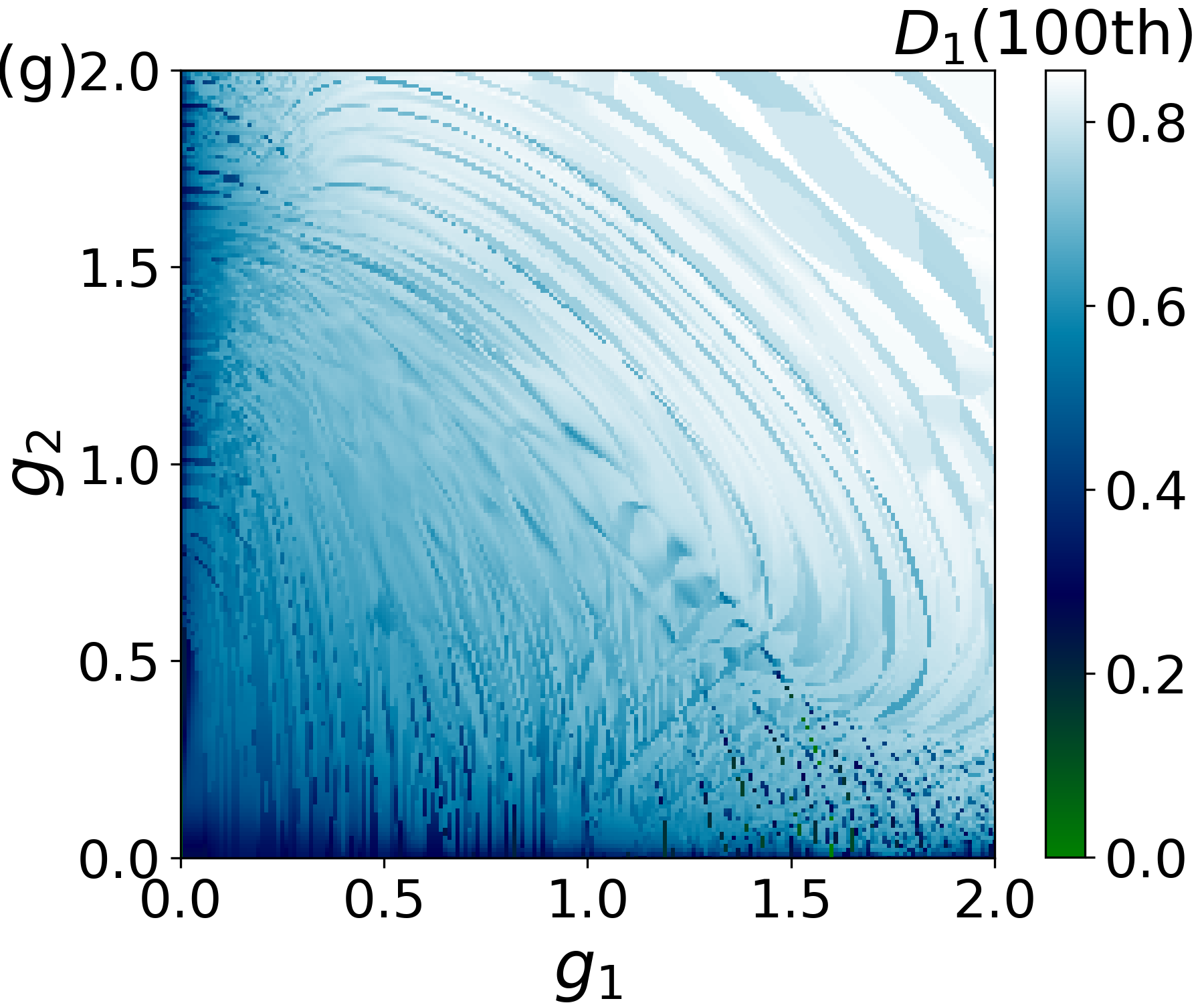}\label{fig:D1_100th}}
      \subfigure{\includegraphics[width=0.245\textwidth]{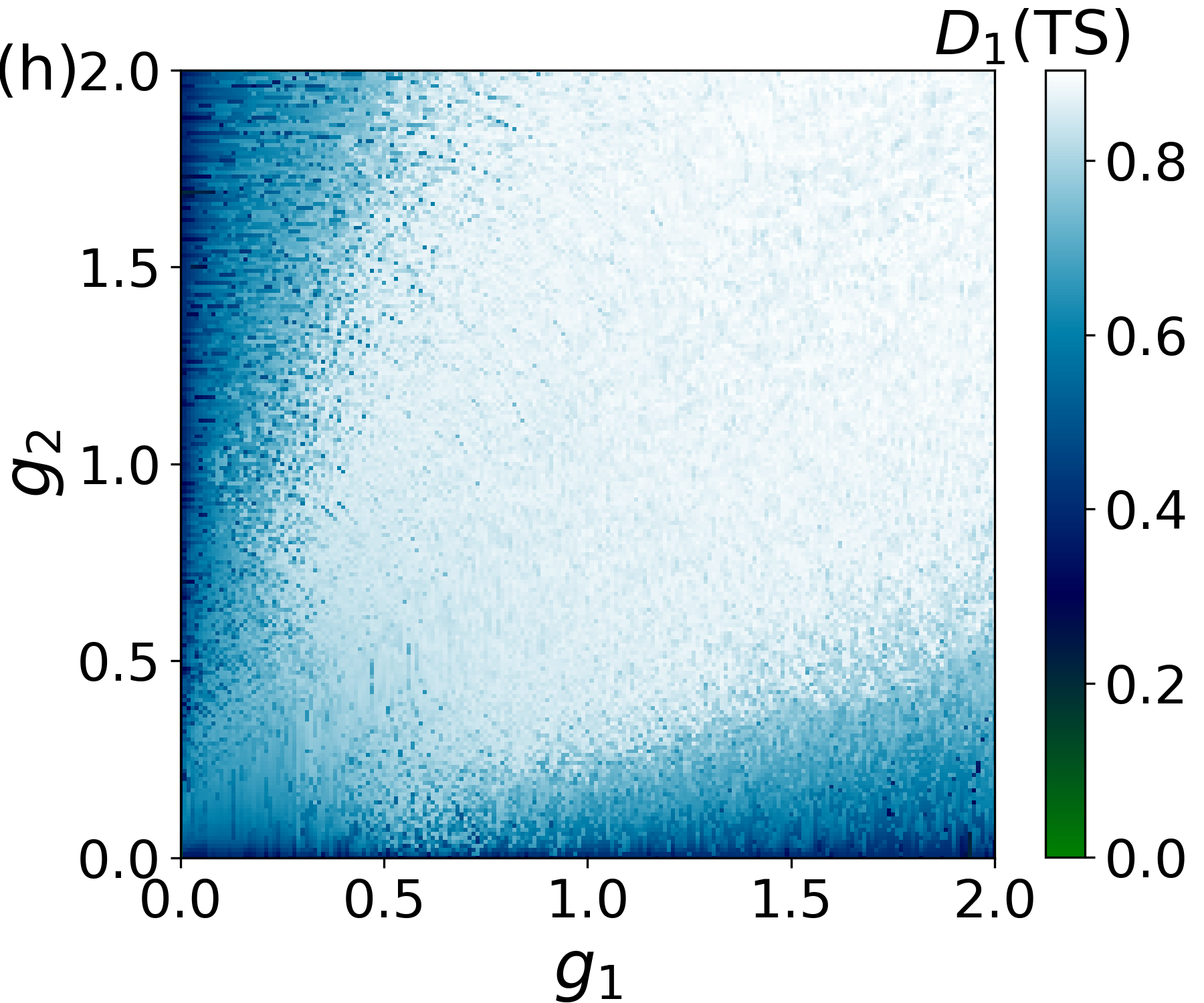}\label{fig:D1_TS}}
      \caption{Participation ratio and multifractal dimension, $D_1$
        of ADM for different eigenstate of the system Hamiltonian as a
        function of coupling parameters $g_1$ and $g_2$. (a), (b),
        (c), (d) are the figure for participation ratio corresponding
        to ground state (gs), tenth excited state, $100{\text{th}}$ excited
        state, and middle excited state or thermal state (TS) respectively. 
        (e) $D_1$ for the ground state shows QPT from localization (NP) 
        to multifractal (SP) behavior, (f) $D_1$ for the tenth
        excited state, (g) $D_1$ for the $100{\text{th}}$ excited state, (h)
        $D_1$ for the middle excited state shows ENET from non-ergodic
        extended (multifractal) to ergodic (delocalized)
        behavior. For all the above plots parameters are: $\omega =
        \omega_0 = 1$, atom number $N=40$. We take the bosonic cut-off
        to be $n_{\text{max}}=200$.}
      \label{fig:enet}
    \end{figure*}
Figure~\ref{fig:r_avg} shows the phase diagram based on the level
spacing ratio. It can be seen that a small value of either $g_1$ or
$g_2$ leads to a non-egrodic phase where the eigenvalue statistics
obey the Poisson distribution with $\langle r\rangle \approx
0.386$. On the other hand when both the couplings are significantly
large, we see a crossover to Wigner-Dyson distribution where the
level-spacing ratio becomes $\langle r\rangle \approx 0.53$. It is
worth mentioning that this non-ergodic to ergodic transition is quite
different from the normal to super-radiant phase transition. We
perform a careful analysis of this phase transition in the next
sub-section. We show that the Wigner-Dyson behavior corresponds to
the energy band that lies between the two
cut-offs~\cite{das2022revisiting} ($E_{\text{lower}} =
E_{\text{normal}}^{0}$, $E_{\text{upper}} =
E_{g_1,g_2=0}^{\text{max}}$).


  So far we restricted ourselves to the resonant case $\omega =
  \omega_0$.  Now we also study ESQPT for the off-resonant cases
  considering $\omega = \frac{\omega_0}{4}$ (middle panel of
  Fig.~\ref{fig:esqpt}) and $\omega = 4\omega_0$ (bottom panel of
  Fig.~\ref{fig:esqpt}). Studying the consecutive level spacing ratio
  of the energy band sandwiched between $E_{\text{normal}}^{0}$ and
  $E_{g_1,g_2= 0}^{\text{max}}$ [see Figs.~\ref{fig:esqpt}(h), and \ref{fig:esqpt}(l)], 
  we see a transition from $\langle r\rangle\approx 0.386$ to
  $\langle r\rangle\approx 0.53$ that is a non-ergodic to ergodic
  transition. From a careful observation of Figs.~\ref{fig:esqpt}(f), and
  \ref{fig:esqpt}(g) and \ref{fig:esqpt}(j), and \ref{fig:esqpt}(k) we infer that while a clear correspondence between
  ENET and ESQPT is present for $g_1\neq g_2$ even in the off-resonant
  scenario, the diagonal direction ($g_1 = g_2$) corresponding to the
  Dicke model exhibits special behavior~\cite{chavez2016classical}.

    \subsection{Ergodic to non-ergodic transition (ENET)} 
    \subsubsection{Statics}
    As discussed in the previous subsection, different eigenstates
    play a role in different phase transitions.  While the ground
    state shows the normal to super-radiant phase transition or
    equivalently from a localized phase to a multifractal phase, the
    middle excited states exhibit a non-ergodic to ergodic
    transition.  Here, we study the phase diagram on the $g_1,\ g_2$
    plane of the ADM for different eigenstates with the help of
    participation ratio and multifractal dimension and explore the 
    possibility of multifractal states in the excited states.
    
    While the participation ratio quantifies the degree of
    localization and delocalization of a quantum state, a study of its
    scaling with the system size offers further insights. When the
    Hilbert space dimension is large ($N_D$ is large) the multifractal
    dimension~ \cite{mace2019multifractal, lindinger2019many} can be
    represented as:
    \begin{eqnarray}
    D_q &=& \frac{1}{1-q}\frac{\ln\Big(\sum_{j=1}^{N_D}\vert\psi_j\vert^{2q}\Big)}{\ln(N_D)}
    \end{eqnarray}
    where $\vert\psi\rangle$ is an eigenstate of the Hamiltonian and
    $S_q =
    \frac{1}{1-q}\ln\Big(\sum_{j=1}^{N_D}\vert\psi_j\vert^{2q}\Big)$
    is known as the $q$-dependent participation entropy. In the
    Shannon limit, $S_1 = - \sum_j\vert\psi_j\vert^{2}\ln\Big(\vert
    \psi_j\vert^{2}\Big)$, while the $q=2$ participation entropy is
    connected to the usual participation ratio as $S_2=\ln(P)$. For a
    perfectly delocalized state $S_q=\ln(N_D)$, (when $N_D$ is large)
    and hence $D_q=O(1)$, for all $q$. On the other hand for
    a localized state $S_q=$ constant and $D_q=0$. In an intermediate
    situation, wave functions are extended but non-ergodic with
    $S_q=D_q\ln(N_D)$ where $0<D_q<1$ and the state is multifractal in
    that particular basis.

    \begin{figure*}[htbp]
      \subfigure{\includegraphics[width=0.248\textwidth]{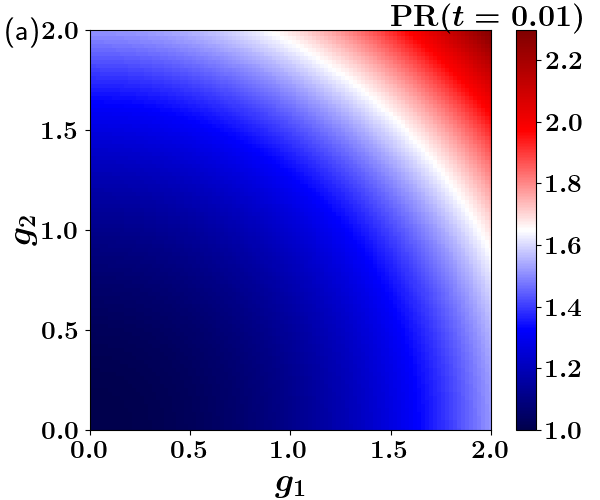}\label{fig:PR_dyna1}}
      \subfigure{\includegraphics[width=0.248\textwidth]{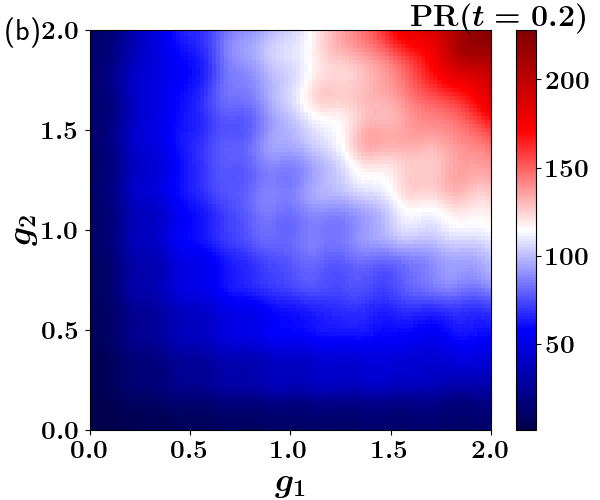}\label{fig:PR_dyna2}}
      \subfigure{\includegraphics[width=0.243\textwidth]{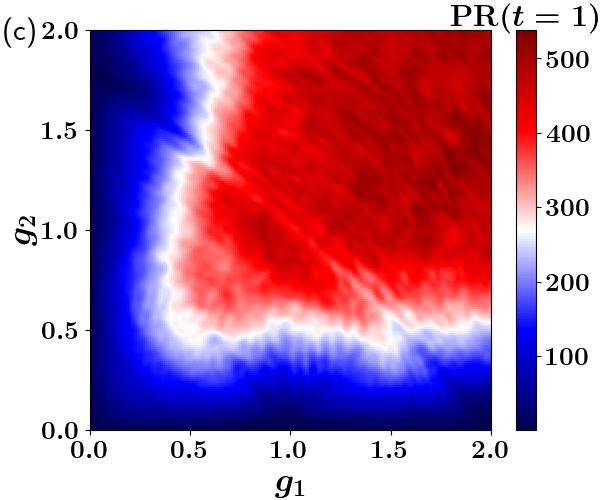}\label{fig:PR_dyna3}}
      \subfigure{\includegraphics[width=0.243\textwidth]{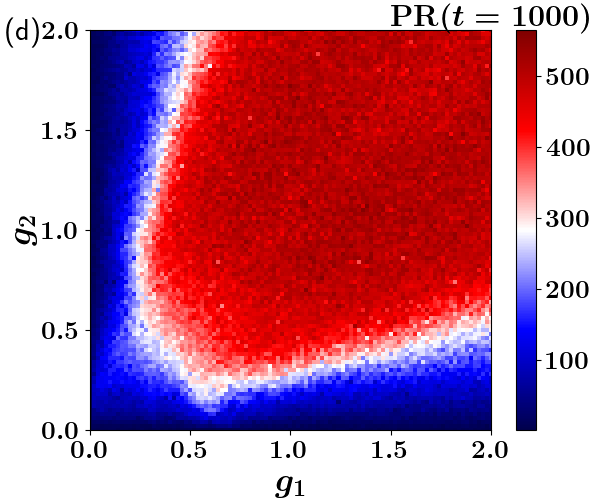}\label{fig:PR_dyna4}}
      \caption{ (a) Quench dynamics of participation ratio of ADM for
        $\omega = \omega_0 = 1$, $N=20$ and $n_{\text{max}}=100$, at
        different times: (a) $t=0.01$, (b) $t=0.2$, (c) $t=1$, (d)
        $t=1000$. Here the initial state is the middle excited state
        of the decoupled Hamiltonian of the system ($H_0 = \omega
        a^{\dagger}a + \omega_{0}J_{z}$). The time $t$
          is in units of $\omega_0^{-1}$ and we fix $\omega_{0}=1$
          throughout this paper.}
      \label{fig:qd1}
    \end{figure*}
        
    In Fig.~\ref{fig:enet} we show the phase diagrams of the ADM based
    on participation ratio [Figs.~\ref{fig:enet}(a) to \ref{fig:enet}(d)] and the
    multifractal dimension $D_1$
    [Fig.~\ref{fig:enet}(e) to \ref{fig:enet}(h)] for different eigenstates
    of the system Hamiltonian.  For a fixed system size, we pick a few
    states including the ground state and the middle excited
    state. While for the ground state, we clearly see the normal to
    super-radiant phase transition along the line $g_1 + g_2 = 1$ as
    shown by the phase diagram of the participation ratio
    [Fig.~\ref{fig:PR_gs}] and the multifractal dimension
    [Fig.~\ref{fig:D1_GS}], as the states become more and more
    excited, it is the non-ergodic to ergodic transition that is
    highlighted.  Studying the PR of the middle excited state
    [Fig.~\ref{fig:PR4}], we see a similar phase diagram as that of
    the level spacing ratio of the system as a function of $g_1$ and
    $g_2$ [see Fig.~\ref{fig:r_avg}], signifying a transition from the
    non-ergodic phase to the ergodic phase. In the non-ergodic phase
    the PR value is low whereas in the ergodic phase its value is
    relatively higher.

    In Figs.~\ref{fig:enet}(e) to \ref{fig:enet}(h) we study the multifractal dimension
    $D_1$ for the different eigenstates of the ADM as a function of
    coupling parameters $g_1$ and $g_2$ similar to the participation
    ratio.  Here in Figure~\ref{fig:D1_GS} we show the nature of $D_1$,
    for the ADM ground state. In the NP, $D_q \approx 0$ whereas in
    the SP, $0<D_1<1$. This suggests a transition from a localized to
    a multifractal phase. Fig.~ \ref{fig:D1_TS} shows $D_1$ for the
    middle excited state of the ADM. In this figure we see the regions
    (depicted by the blue color) where $0<D_1<1$ which behaves like a
    multifractal phase, whereas the region where $D_1\approx 0.9$
    (depicted by the white color), behaves as more like a delocalized
    phase.  Hence the middle excited state shows a transition from an
    extended non-ergodic (multifractal) phase to an ergodic
    (delocalized) phase.

    \subsubsection{Dynamics}
    	  To study the quench dynamics of a closed quantum
          system, one prepares the system in some eigenstate of the
          initial Hamiltonian $\mathcal{H}_0$. The Hamiltonian is
          suddenly changed to
          $\mathcal{H}=\mathcal{H}_0+\mathcal{H}_1$, and the system is
          allowed to evolve under the corresponding unitary time
          evolution operator. Here, we take the middle excited state
          of the decoupled Hamiltonian ($\mathcal{H}_0 = \omega
          a^{\dagger}a + \omega_0 J_z$), as our initial state which
          can be written as $|\psi_{\text{in}}\rangle = \sum_\alpha
          C_\alpha|\alpha\rangle$, with $|\alpha\rangle =
          |n,j,m\rangle$ being a computational basis state and
          $C_\alpha$ being the corresponding coefficient. The
          time-evolution of the state is given by $|\psi_t\rangle =
          e^{-i\mathcal{H}t}| \psi_{\text{in}}\rangle = \sum_\alpha
          C_\alpha(t)|\alpha\rangle$, where $\mathcal{H}$ is the ADM
          Hamiltonian. To study the dynamical properties, we calculate
          the dynamics of the participation ratio $ \text{PR}(t) =
          1/{\sum_{\alpha} |C_\alpha(t)|^4}$, at different times:
          $t=0.01$, $0.2$, $1$, $1000$.  From Fig.~\ref{fig:PR_dyna1}
        we see that, for a very small duration of time say $t=0.01$,
        the participation ratio has low value for all $g_1$ and $g_2$.
        As we evolve the system a bit,
        say at $t=0.2$, in Fig.~\ref{fig:PR_dyna2}, one can notice
        that in the red part, the PR value is increasing, for higher
        values of $g_1$ and $g_2$. In fact this portion in the central
        region increases with time also. In Fig.~\ref{fig:PR_dyna3} we
        see that in the red part the value of PR is significantly
        higher than in the blue part. When the dynamics is carried out
        over a very long time [say $t= 1000$ as in
        Fig.~\ref{fig:PR_dyna4}], we get a phase diagram which exactly
        looks like the non-ergodic (blue color) to ergodic (red color)
        phase diagram [see Fig.~\ref{fig:PR_dyna4} and
          \ref{fig:PR4}].  Figure~\ref{fig:PR_dyna4} suggests that,
        if the system is initially in the non-ergodic phase, and we
        evolve it for a long time, the system will stay in the
        non-ergodic phase, i.e., the participation ratio is relatively
        low no matter how long the time is. This is in contrast to the
        ergodic phase where the PR value is higher for long times even
        if the inital PR is low.

    \section{Thermal phase transition (TPT)}\label{sec_4}
    \begin{figure}[t]
      \includegraphics[width=0.45\textwidth]{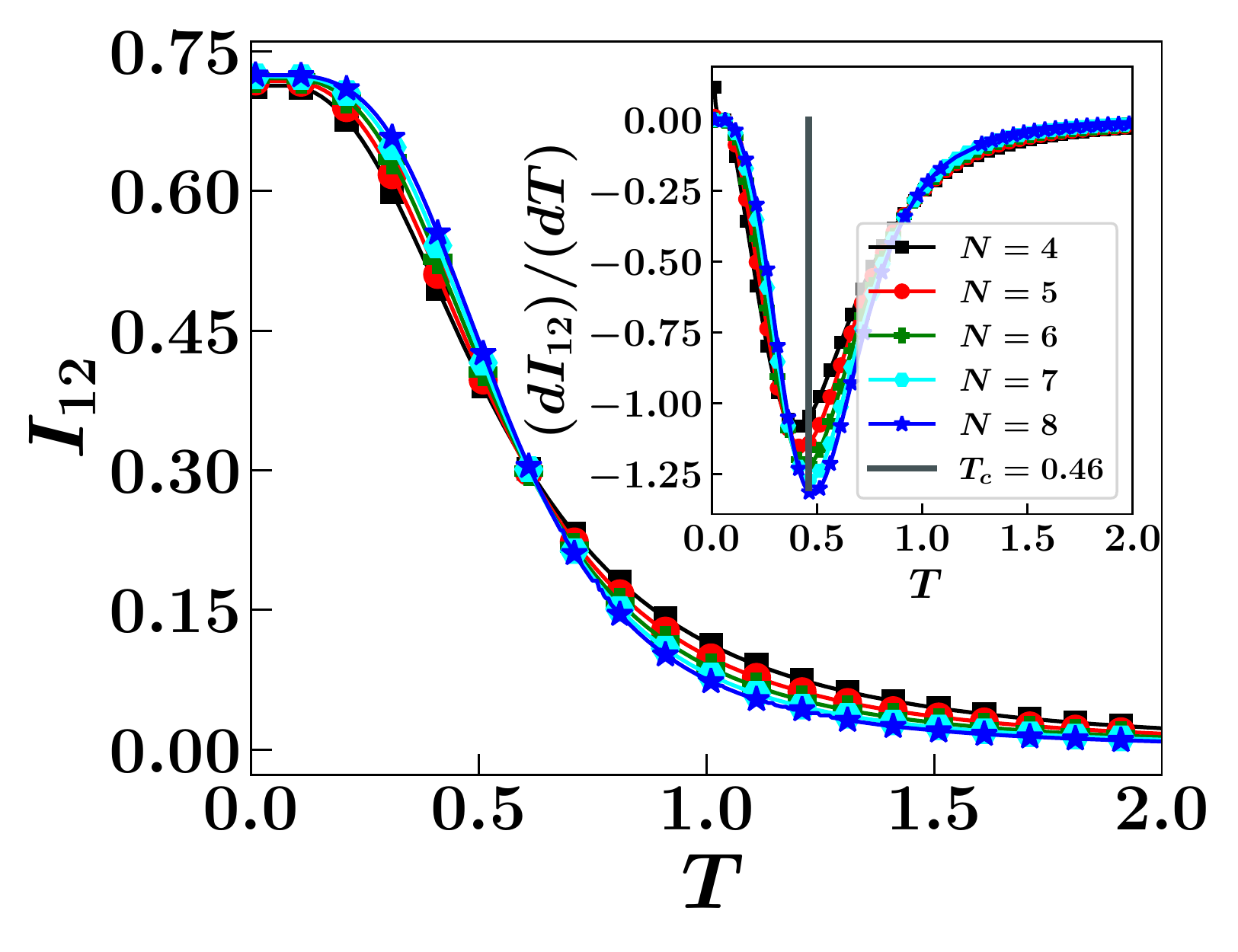}
      \caption{Mutual information of two spins $I_{12}$ as a function of temperature. Inset of the figure shows the numerical differentiation of MI with respect to temperature $\frac{dI_{12}}{dT}$, at $g_1 = 1.0$, $g_2 = 0.5$. The vertical line represents the theoretical value of the critical temperature, $T_c\approx 0.46$. Here $\omega = \omega_0 = 1$, $n_{\text{max}}=40$.}
      \label{fig:tpt1}
    \end{figure}
    
    \begin{figure*}[htbp]
      \subfigure{\includegraphics[width=0.24\textwidth]{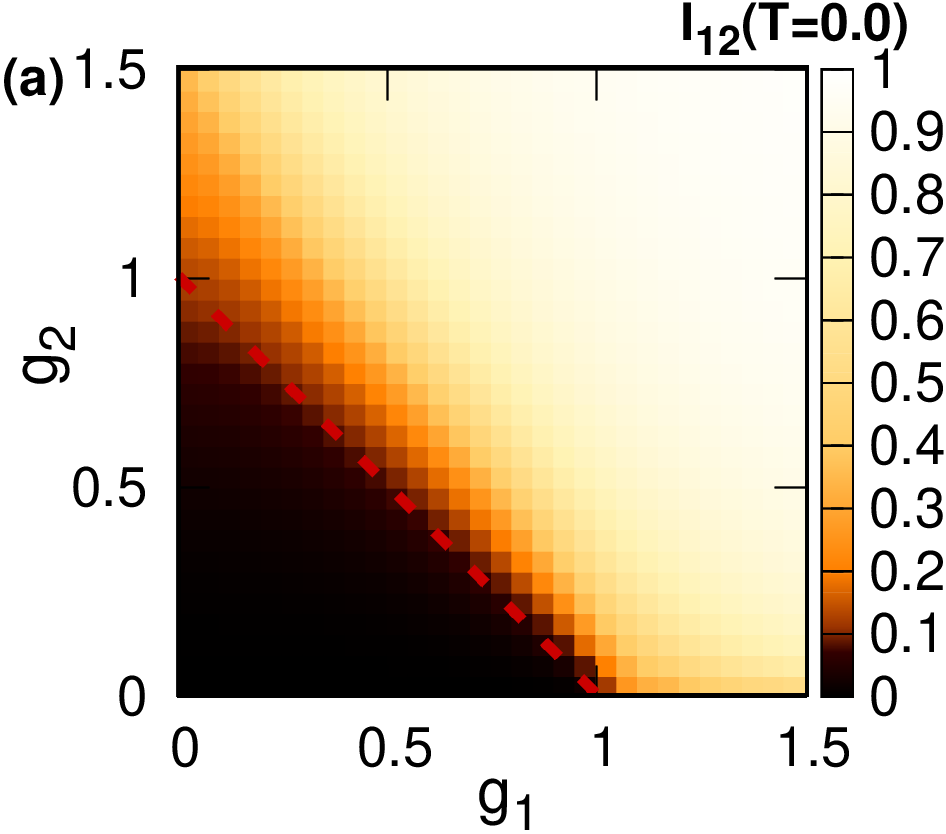}\label{fig:MI_T0pt0}}
      \subfigure{\includegraphics[width=0.24\textwidth]{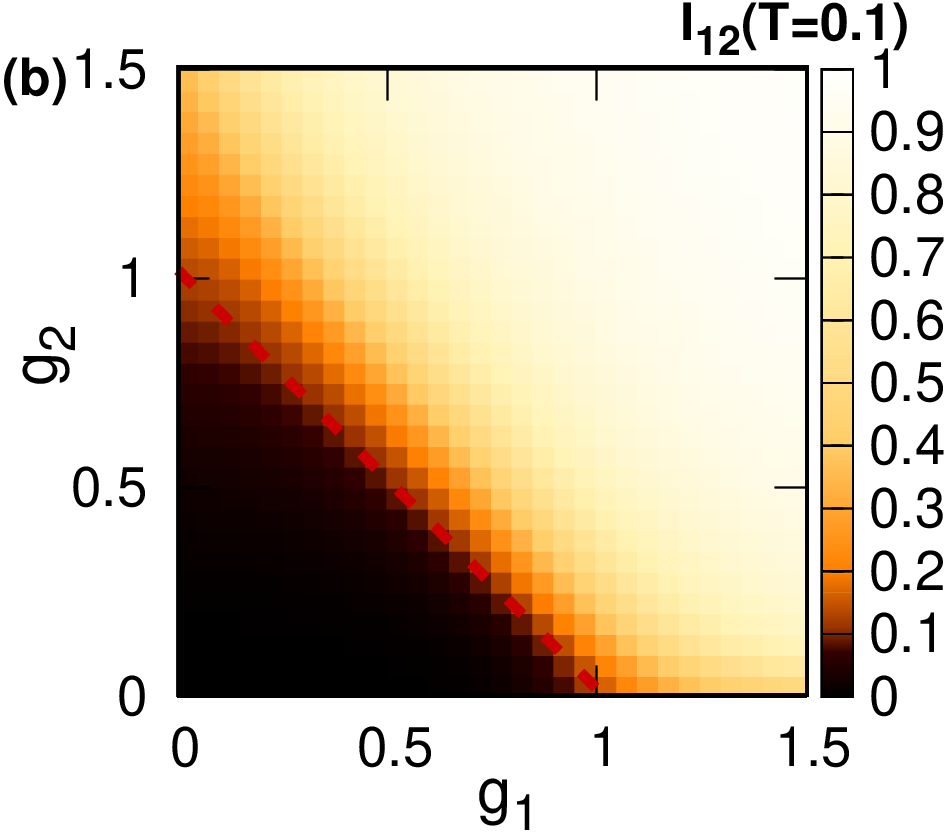}\label{fig:MI_T0pt1}}
      \subfigure{\includegraphics[width=0.24\textwidth]{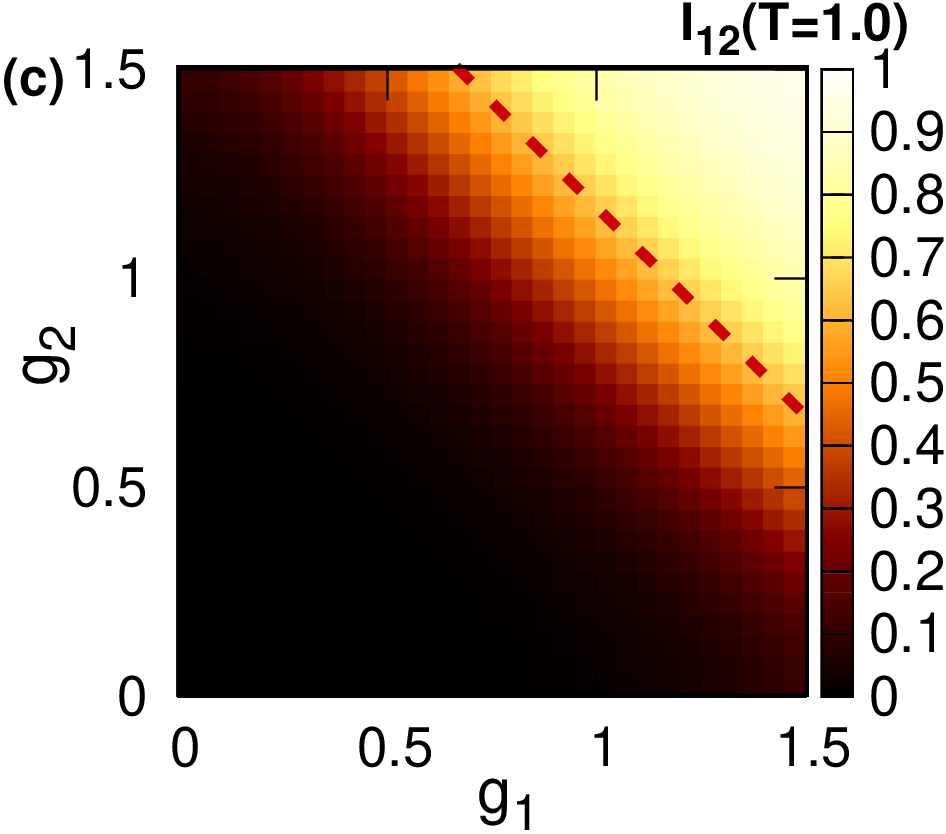}\label{fig:MI_T1pt0}}
      \subfigure{\includegraphics[width=0.25\textwidth]{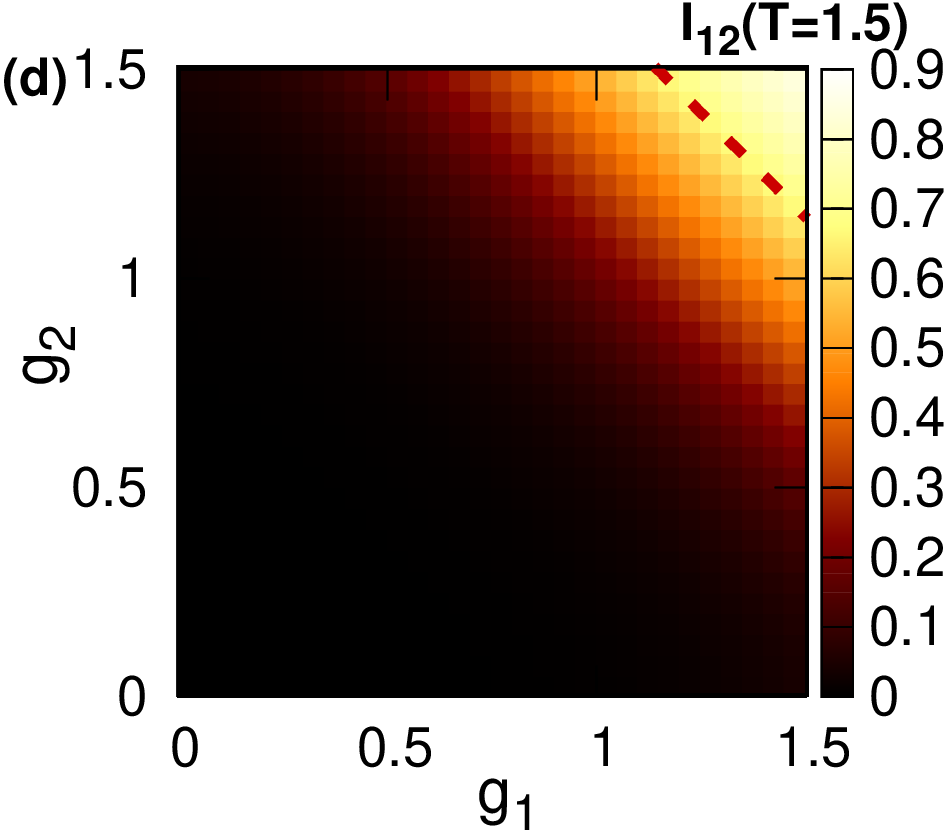}\label{fig:MI_T1pt5}}
      \caption{Mutual information between two spins as a function of
        $g_1$ and $g_2$ at different temperature: (a) $T=0$ (QPT), (b)
        $T=0.1$, (c) $T=1.0$, (d) $T=1.5$. As temperature increases,
        the region corresponding to the normal phase (denoted by black
        color) also increases. The red dashed line denotes the value
        of $g_1+g_2$ for a fixed temperature following
        Eq.~\ref{eqn:Tc}. Zero temperature case denotes QPT along the
        line $g_1+g_2=1$, for $T=0.1$, $1$ and $1.5$, $g_1+g_2=1.01$,
        $2.16$, $2.63$ respectively. The parameters are: $\omega =
        \omega_0 = 1$, $N=6$. The bosonic cut-off is taken to be
        $n_{\text{max}}=40$.}
      \label{fig:tpt2}
    \end{figure*}
    Another type of phase transition exhibited by the ADM is the
    finite temperature phase transition~\cite{hioe1973phase}. As with
    the Dicke model, it is known that a finite critical temperature
    can destroy the super-radiant phase and a transition back to the
    normal phase is obtained going beyond the critical
    temperature~\cite{hioe1973phase}. We can derive an exact
    analytical expression for the transition temperature as a function
    of the parameters $g_1$ and $g_2$. We start by rewriting the
    Hamiltonian (in units of $\omega$) as:
    \begin{align}
    \tilde{\mathcal{H}} = \frac{\mathcal{H}}{\omega} = a^{\dagger}a + \sum_{j=1}^{N}\frac{\epsilon}{2}\sigma_j^z + \frac{\lambda_1}{2\sqrt{N}}\sum_{j=1}^{N}( a\sigma_j^+ + a^{\dagger}\sigma_j^- )\nonumber\\
    + \frac{\lambda_2}{2\sqrt{N}}\sum_{j=1}^{N}( a^{\dagger}\sigma_j^+ + a\sigma_j^- )
     \end{align}
     where, $\epsilon = \frac{\omega_0}{\omega}$, $\lambda_1=\frac{g_1}{\omega}$, $\lambda_2=\frac{g_2}{\omega}$.
     The partition function for the full ADM is given by:
     \begin{align}
      Z(N,T) = \sum_{s_1, . . . , s_N=\pm 1}\int \frac{d^2\alpha}{\pi}\langle s_1, . . .,s_N\vert\langle\alpha\vert e^{-\beta\tilde{\mathcal{H}}}\vert\alpha\rangle\vert s_1, . . . , s_N\rangle.
    \end{align}
    The expectation value of the Hamiltonian with respect to the
    bosonic modes is:
     \begin{align}
     \langle\alpha\vert \tilde{\mathcal{H}}\vert\alpha\rangle = \alpha^*\alpha + \sum_{j=1}^N\Big[ \frac{\epsilon}{2}\sigma_j^z + \frac{\lambda_1}{2\sqrt{N}}( \alpha\sigma_j^+ + \alpha^{*}\sigma_j^- )\nonumber\\
      + \frac{\lambda_2}{2\sqrt{N}}( \alpha^{*}\sigma_j^+ + \alpha\sigma_j^- ) \Big].
     \end{align}
     Defining
     \begin{align}
     h_j = \frac{\epsilon}{2}\sigma_j^z + \frac{\lambda_1}{2\sqrt{N}}( \alpha\sigma_j^+ + \alpha^{*}\sigma_j^- )
      + \frac{\lambda_2}{2\sqrt{N}}( \alpha^{*}\sigma_j^+ + \alpha\sigma_j^- )
     \end{align}
     the expectation value with respect to the spins becomes a product
    of single-spin expectation values:
     \begin{align}
     \langle s_1...s_N\vert\langle\alpha\vert e^{-\beta \tilde{\mathcal{H}}}\vert\alpha\rangle\vert s_1...s_N\rangle &=& e^{-\beta\vert\alpha\vert^2}\Pi_{j=1}^N\langle s_j\vert e^{-\beta h_j}\vert s_j\rangle.
     \end{align}
     Thus the computation of the partition function reduces to the evaluation of a double integral:
     \begin{align}
     Z(N,T) = \int\frac{d^2\alpha}{\pi}e^{-\beta\vert\alpha\vert^2}\Big[ \text{Tr}e^{-\beta h} \Big]^N\nonumber\\
     = \int\frac{d^2\alpha}{\pi}e^{-\beta\vert\alpha\vert^2}\Big( 2\cosh\Big[ \frac{\beta\epsilon}{2}
      \Big[ 1 + \frac{4(\lambda_1+\lambda_2)^2\alpha^2}{\epsilon^2 N} \Big]^{1/2} \Big] \Big)^N  
     \label{eqn:partition_func}   
     \end{align} 
     which in the thermodynamic limit ($N\to\infty$) can be solved
     using the method of steepest descent, within the super-radiant
     phase. Tracking the point at which the method breaks down (see
     the appendix), we have an exact expression for the transition
     temperature:
     \begin{equation}
       T_c = \frac{1}{\beta_c} = \Big(\frac{\omega_0}{2\omega}\Big)\frac{1}{\tanh^{-1}\Big( \frac{\omega\omega_0}{(g_1+g_2)^2} \Big)}.
       \label{eqn:Tc}
     \end{equation}

    Thus, in the super-radiant phase ($g_1+g_2>1$ at $T=0$), raising
    the temperature to a value larger than the critical temperature
    ($T_c$), causes the system to go back to the normal phase.  We now
    provide numerical evidence of this phase transition with the help
    of mutual information between two spins. While the entanglement
    entropy is a good measure to capture a QPT, it is unsuitable for a
    TPT since mixed states are involved. Mutual
    information~\cite{divincenzo2004locking, lu2011optimal,
      henderson2001classical, adesso2010quantum}, between two spins is
    a good measure to capture the TPT, as we showned in an earlier
    work for the Dicke model~\cite{das2022revisiting}. We show here
    that the usefulness of mutual information as a marker of the
    thermal phase transition extends to the ADM.

    When the overall state is mixed, the correlations between two
    subsystems can be quantified with the help of the mutual
    information defined as
    \begin{equation}
    I_{12} =  S_1 + S_2 - S_{12},
    \end{equation} 
    where $S_{1,2} = -\text{Tr}[\rho_{1,2} \ln(\rho_{1,2})]$ , $S_{12}
    = -\text{Tr}[\rho_{12}\ln(\rho_{12})]$. Here $\rho_{1}$,
    $\rho_{2}$ are the reduced density matrices for the two
    subsystems, $S_{1}$, $S_{2}$ are the corresponding von Neumann
    entropies, $\rho_{12}$ is the density matrix of the overall system,
    and $S_{12}$ is the corresponding entropy.  When the overall state
    is in a pure state, $S_{12} = 0$ and the mutual information
    become twice the entanglement entropy since $S_{1}=S_{2}$.  For
    our model we study the mutual information between any two spins -
    (due to the symmetry of the system Hamiltonian, it does not matter
    which spin pair is chosen). Here we use the spin product space
    hence we have to diagonalize the system Hamiltonian with dimension
    $( n_{\text{max}} + 1 )2^N$.  To calculate $I_{\text{12}}$, we
    have to take a partial trace of the total density matrix over the
    bosonic part first and then over the $N-2$ atoms.

    Figure~\ref{fig:tpt1} shows the mutual information between two
    spins. For the parameters $g_1=1.0, g_2=0.5$, the system is in the
    super-radiant phase at zero temperature where the value of mutual
    information takes significantly large values
    (Fig.~\ref{fig:tpt1}). On increasing the temperature, we see that
    the mutual information starts to decrease, signifying a change in
    the direction of the normal phase. To check that mutual
    information does indeed capture the exact transition between the
    super-radiant-to-normal phase, we plot the derivative of the
    mutual information with respect to the temperature: $\frac
    {d{I_{12}}}{dT}$ (inset of the Fig.~\ref{fig:tpt1}). We observe that the
    temperature at which the derivative is minimum corresponds to the
    transition temperature of the TPT from SP to NP for $g_1+g_2>1$.  For
    this particular choice of $g_1$ and $g_2$, the critical
    temperature is $T_c=0.46$, which we denote by the vertical
    straight line in the inset figure.

    In Fig.~\ref{fig:tpt2} we showned $I_{\text{12}}$ as a function
    of $g_1$ and $g_2$ at different temperatures: $T=0$, $0.1$, $1.0$,
    and $1.5$. Figure~\ref{fig:MI_T0pt0} ($T=0$) shows a clear QPT from
    NP (black color) to SP (white color) along the line $g_1+g_2=1$. We
    emphasize that at zero temperature, the ground state is a pure
    state, and so the mutual information is really the same as twice
    the entanglement entropy. $I_{12}$ is close to zero in the normal
    phase and close to unity in the super-radiant phase. Another way
    of saying this is that the total correlation between two spins is
    almost zero in the normal phase whereas it is maximum in the
    super-radiant phase.  Hence the QPT in the anisotropic Dicke model
    is similar to the isotropic Dicke model (for which $g_1=g_2$) but
    here an additional parameter is introduced.  On the other hand,
    one can notice that as temperature increases [in
    Fig.~\ref{fig:MI_T0pt1} ($T=0.1$),  \ref{fig:MI_T1pt0} ($T=1$) and 
    \ref{fig:MI_T1pt5} ($T=1.5$)] the region corresponding to the NP 
    (black portion) also expands - the phase boundary is highlighted by 
    a red dashed line ($g_1+g_2=\text{constant}$) which depends on the
    temperature. Using Eq.~\ref{eqn:Tc} we see that for $T=0.1$, $1$,
    and $1.5$, the phase boundary is given by $g_1+g_2=1.01$,
    $g_1+g_2=2.16$, and $g_1+g_2=2.64$, respectively; beyond
    the phase boundary, the system is in the SP where the correlation
    between two spins is close to unity. These figures indicate that
    the mutual information between spins is an excellent measure of
    the thermal phase transition of the system.

\begin{figure*}[t]
    \subfigure[]{\includegraphics[width=0.18\textwidth]{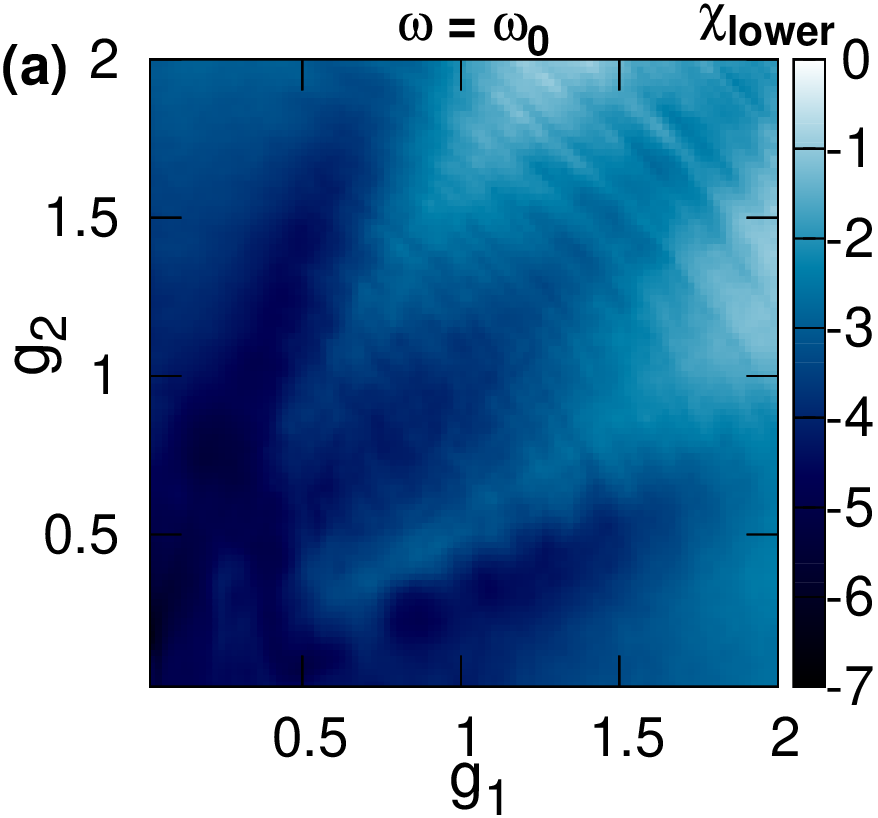}\label{fig:7a}}
    \subfigure[]{\includegraphics[width=0.195\textwidth]{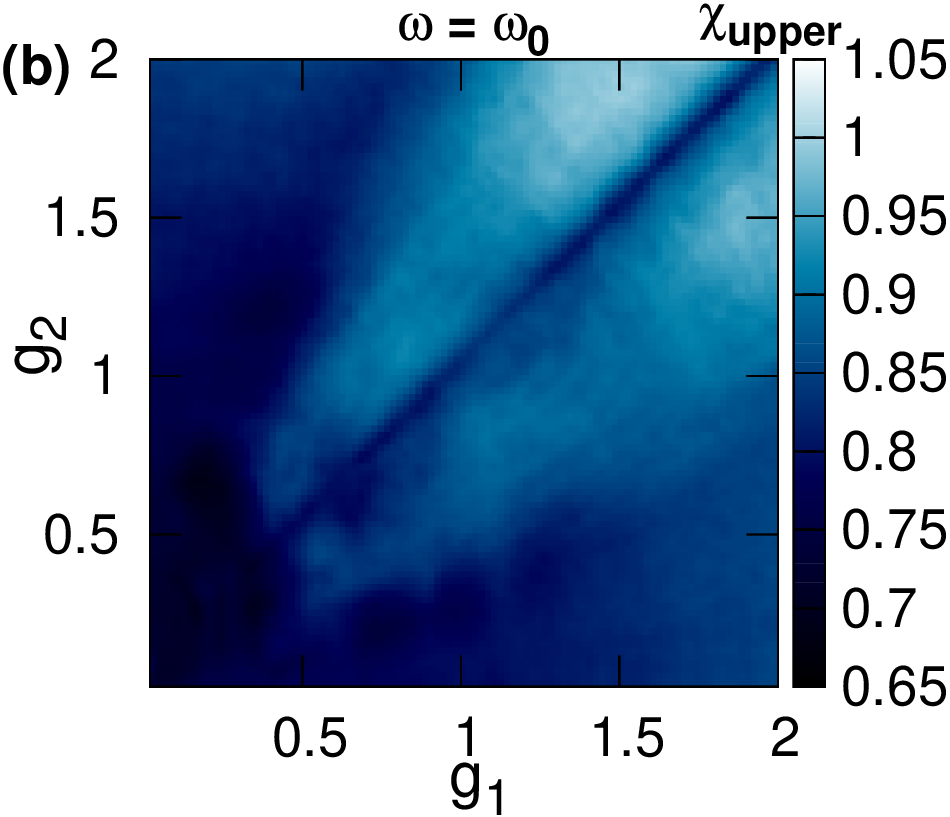}\label{fig:7b}}
    \subfigure[]{\includegraphics[width=0.195\textwidth]{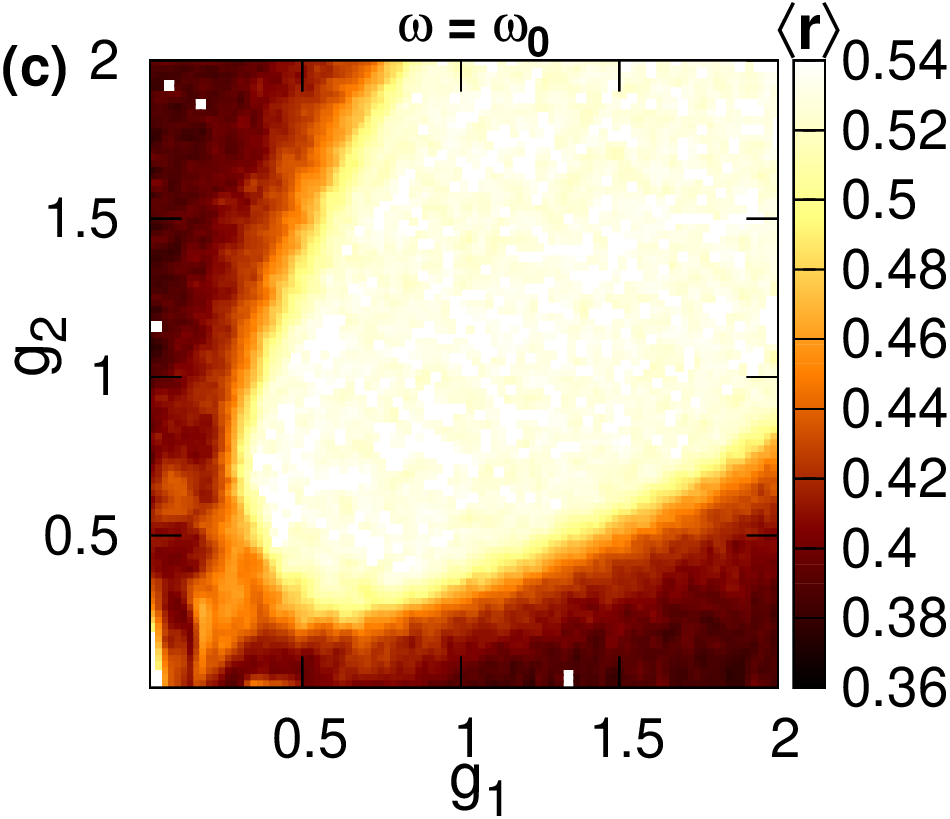}\label{fig:7c}}
	\subfigure[]{\includegraphics[width=0.195\textwidth]{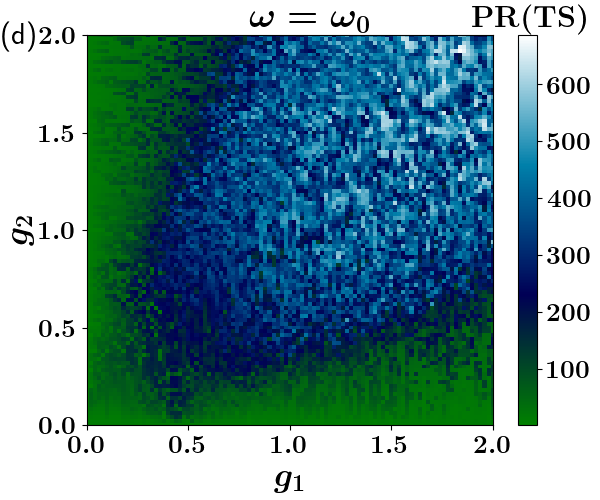}\label{fig:7d}}
	\subfigure[]{\includegraphics[width=0.195\textwidth]{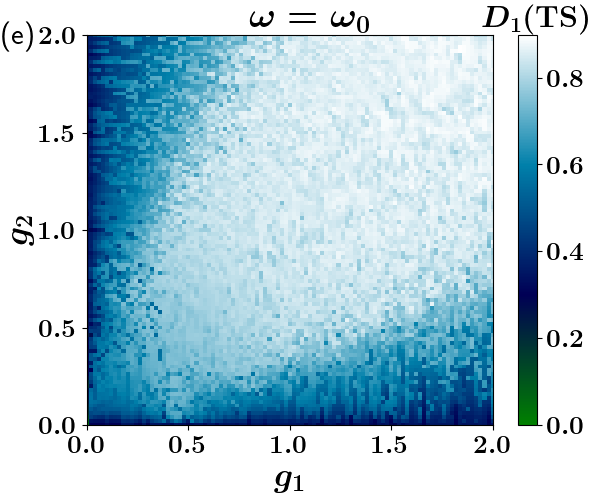}\label{fig:7e}}
	
	\subfigure[]{\includegraphics[width=0.18\textwidth]{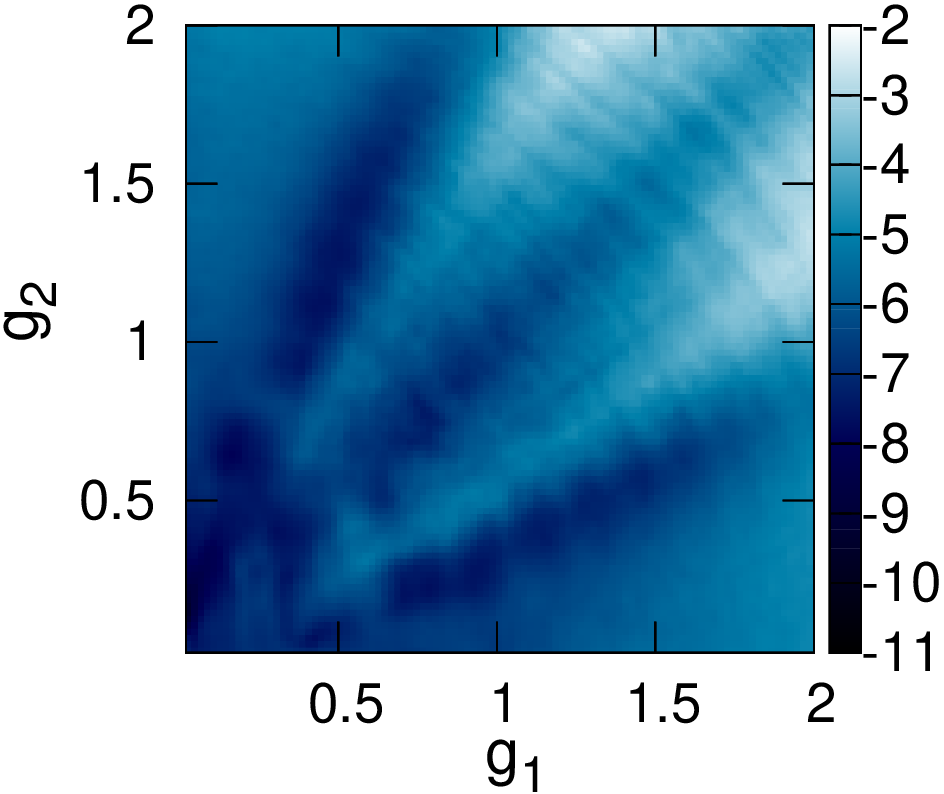}\label{fig:7f}}
    \subfigure[]{\includegraphics[width=0.195\textwidth]{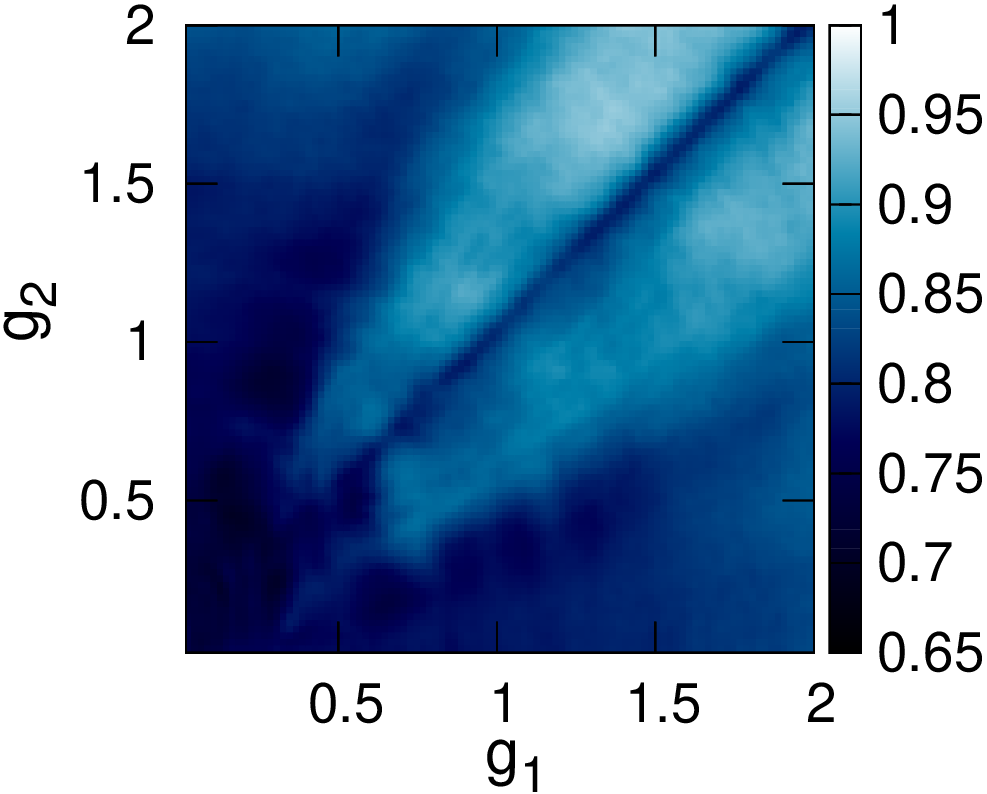}\label{fig:7g}}
    \subfigure[]{\includegraphics[width=0.195\textwidth]{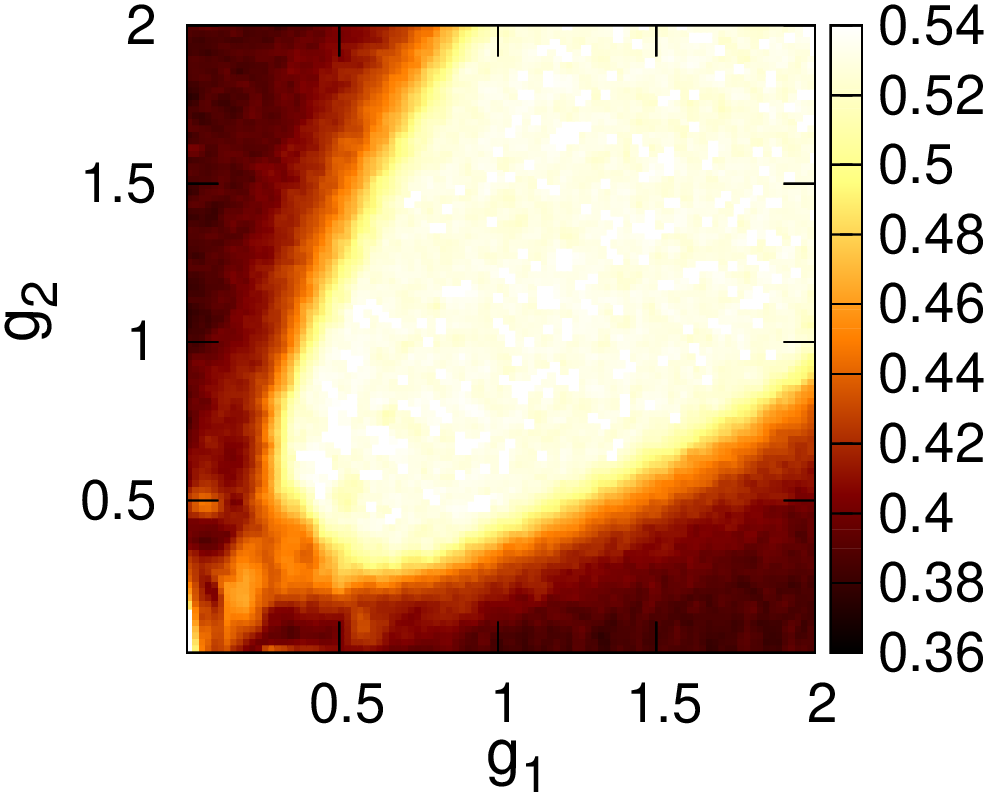}\label{fig:7h}}
	\subfigure[]{\includegraphics[width=0.19\textwidth]{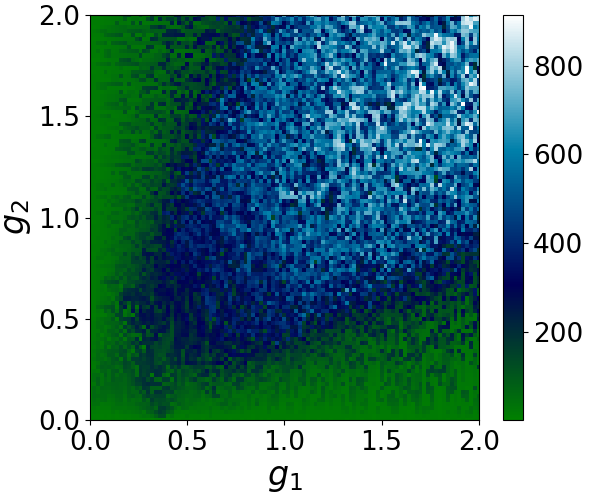}\label{fig:7i}}
	\subfigure[]{\includegraphics[width=0.19\textwidth]{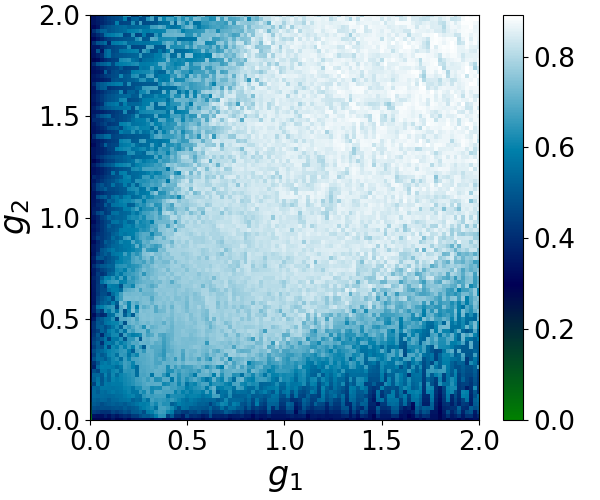}\label{fig:7j}}
	
	\subfigure[]{\includegraphics[width=0.18\textwidth]{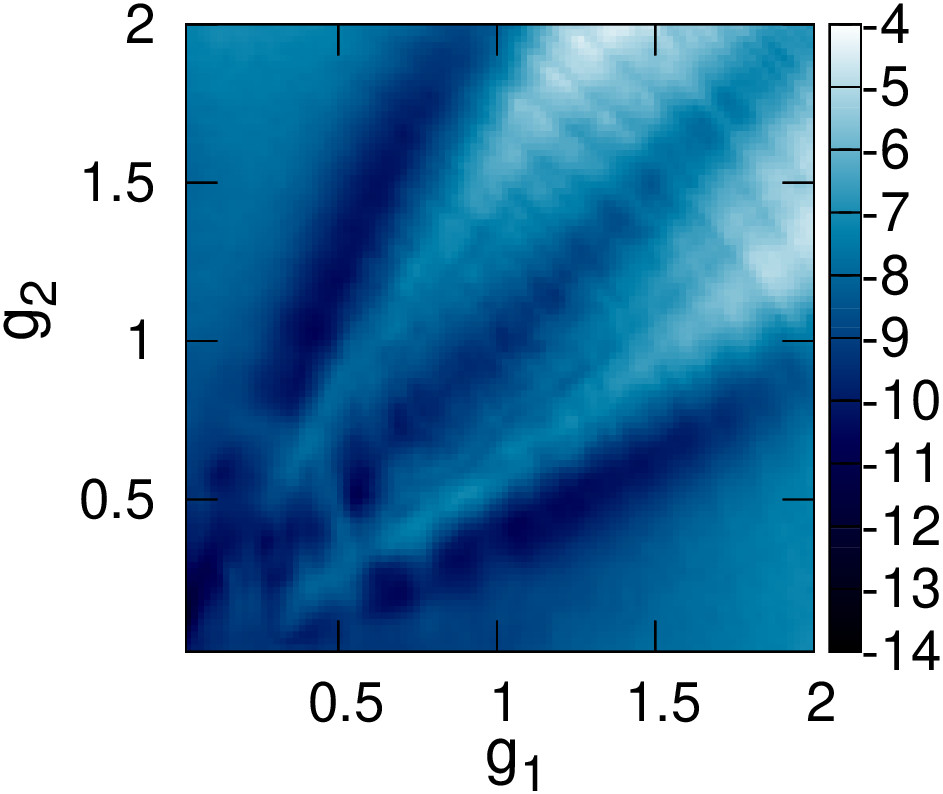}\label{fig:7k}}
    \subfigure[]{\includegraphics[width=0.195\textwidth]{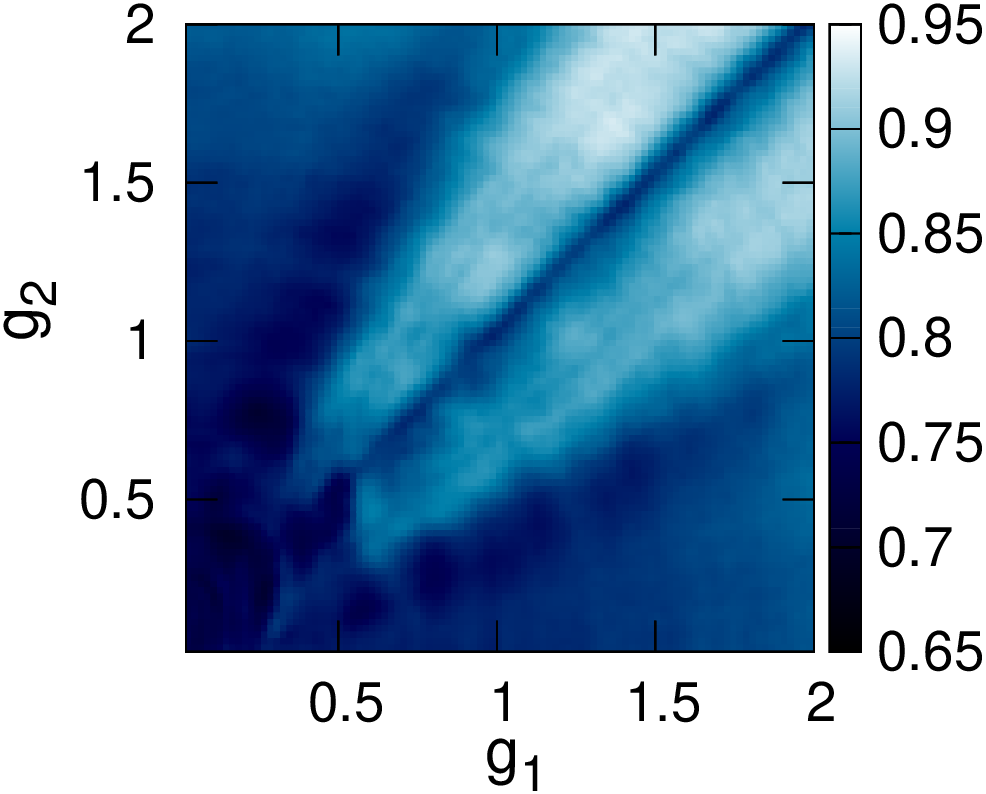}\label{fig:7l}}
    \subfigure[]{\includegraphics[width=0.195\textwidth]{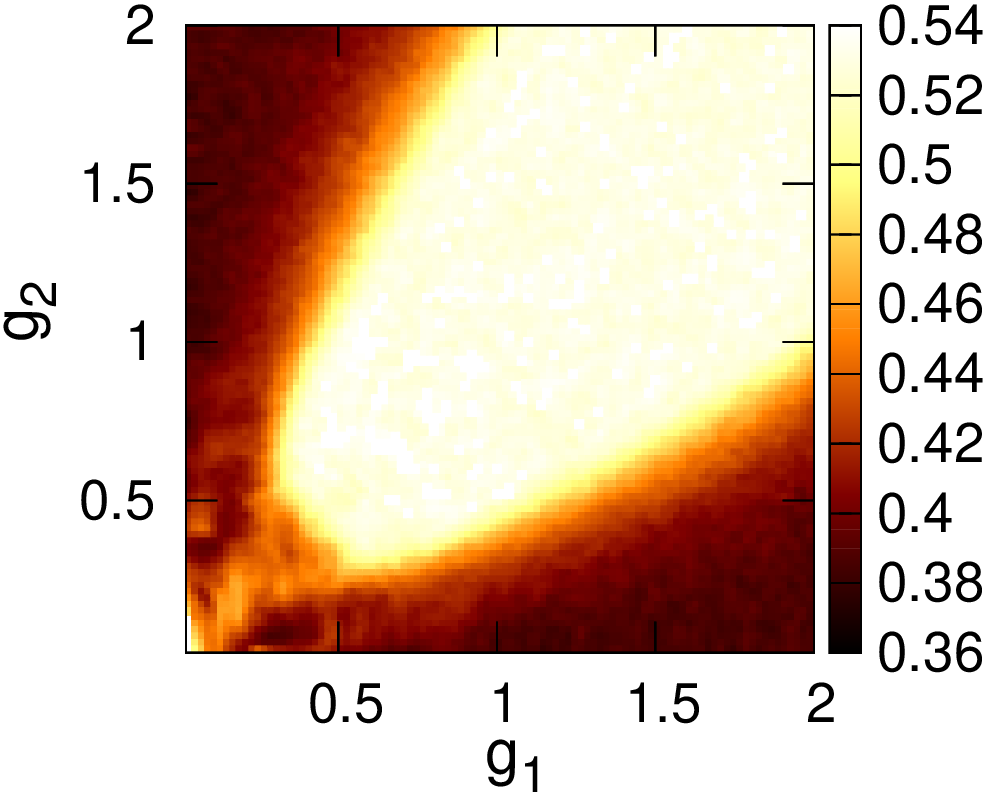}\label{fig:7m}}
	\subfigure[]{\includegraphics[width=0.19\textwidth]{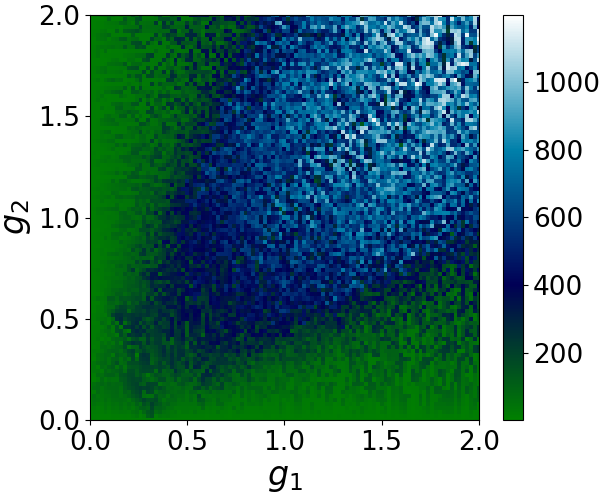}\label{fig:7n}}
	\subfigure[]{\includegraphics[width=0.19\textwidth]{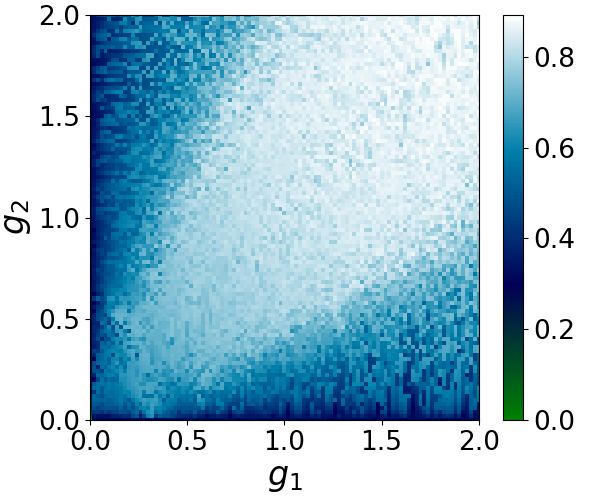}\label{fig:7o}}
	
	\subfigure[]{\includegraphics[width=0.18\textwidth]{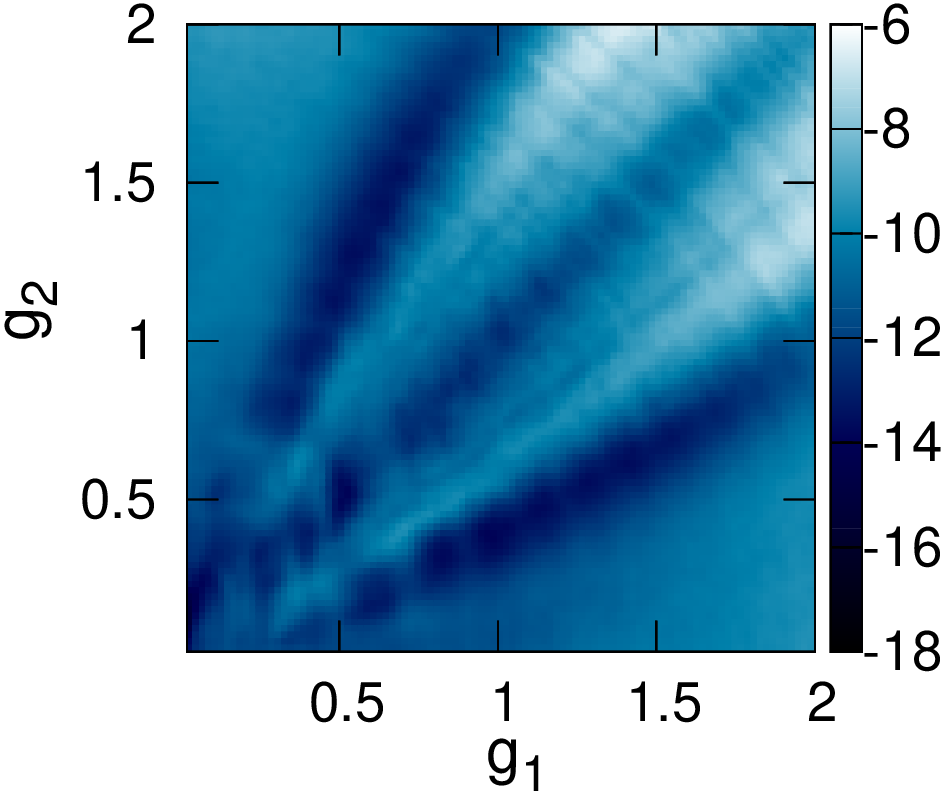}\label{fig:7p}}
    \subfigure[]{\includegraphics[width=0.195\textwidth]{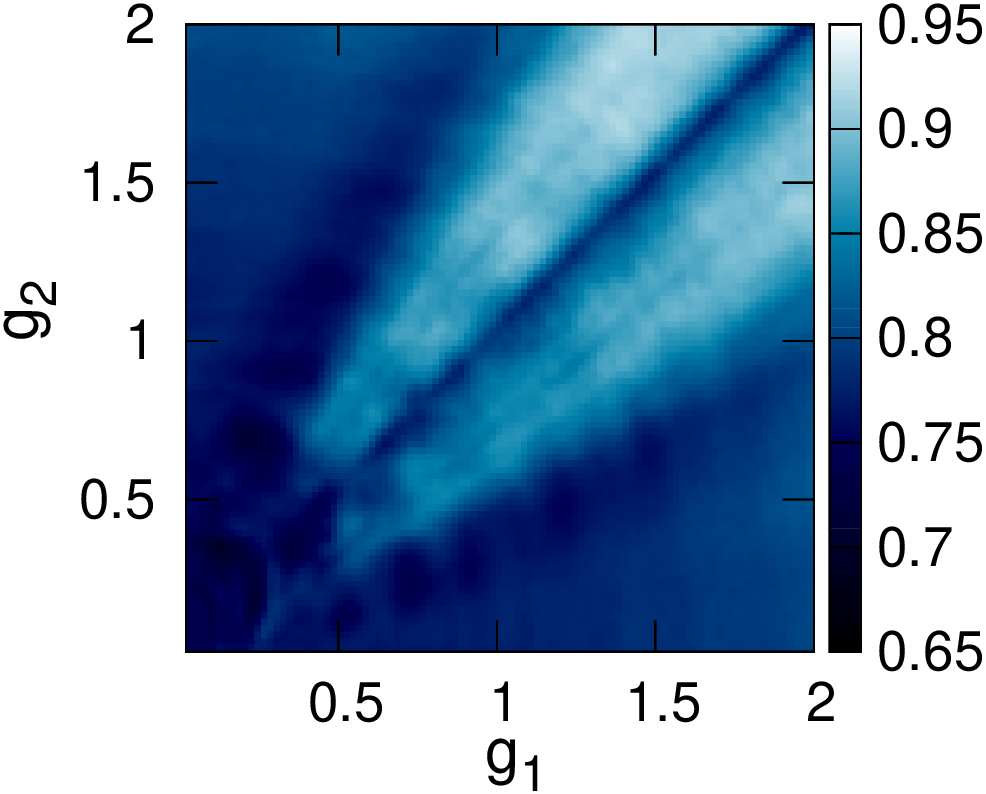}\label{fig:7q}}
    \subfigure[]{\includegraphics[width=0.195\textwidth]{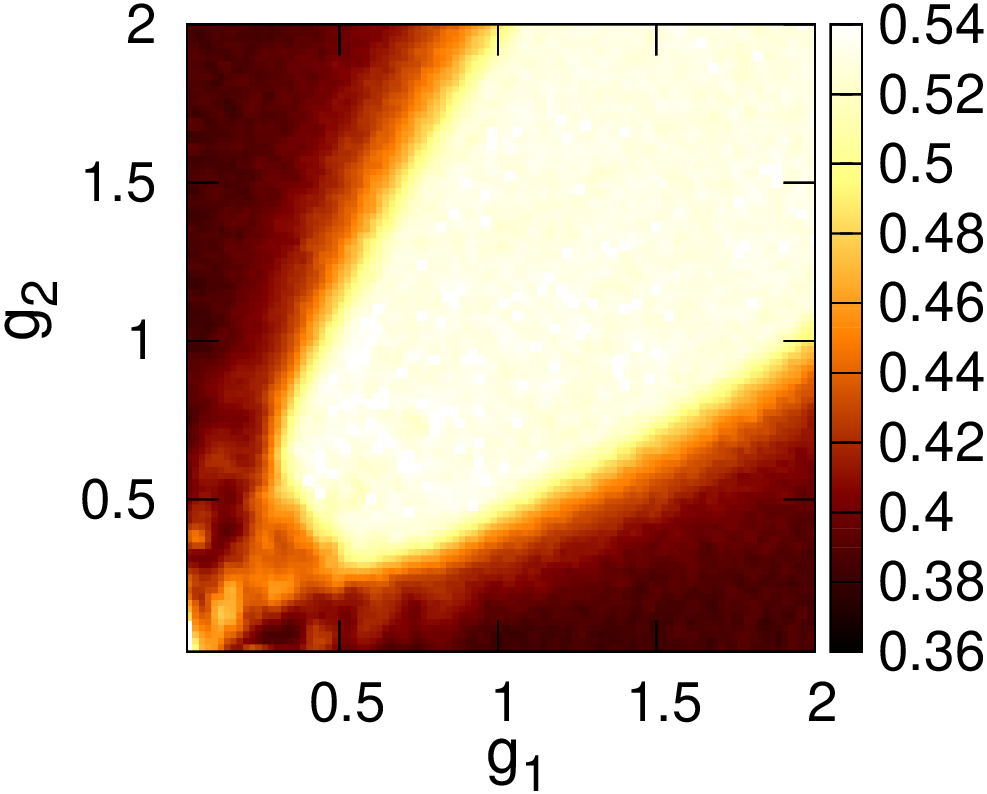}\label{fig:7r}}
	\subfigure[]{\includegraphics[width=0.19\textwidth]{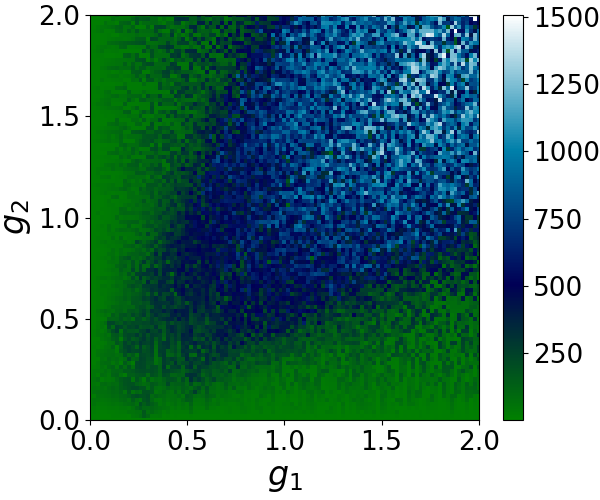}\label{fig:7s}}
	\subfigure[]{\includegraphics[width=0.19\textwidth]{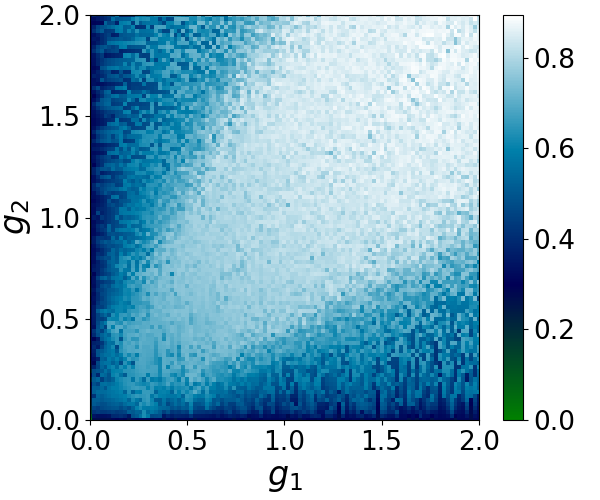}\label{fig:7t}}
	
	\subfigure[]{\includegraphics[width=0.18\textwidth]{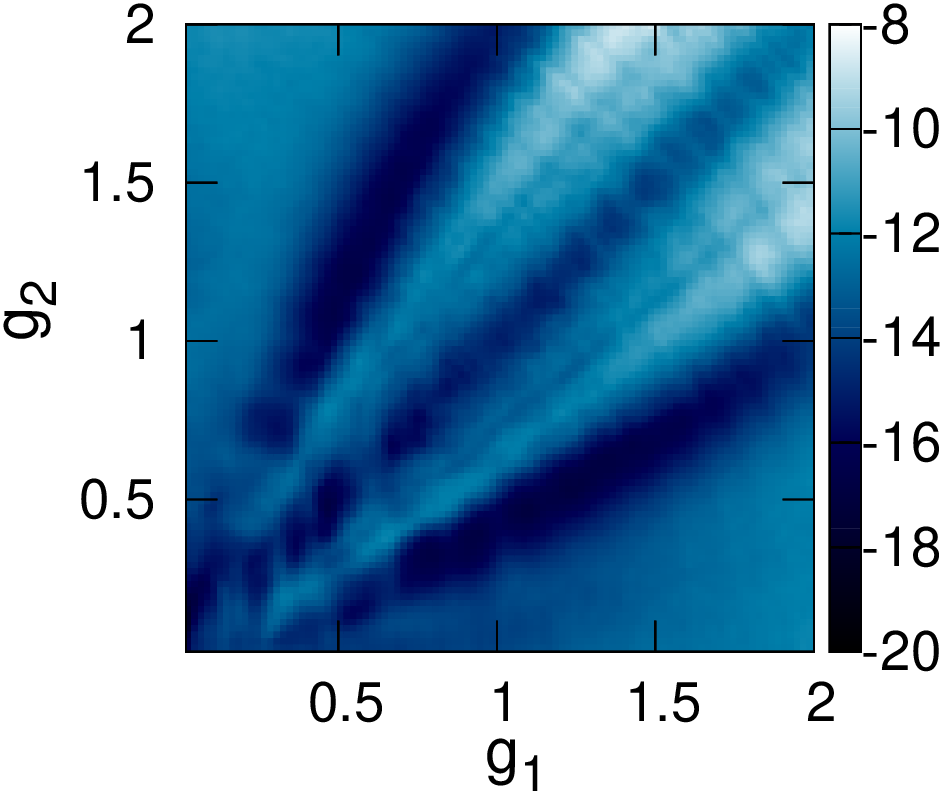}\label{fig:7u}}
    \subfigure[]{\includegraphics[width=0.195\textwidth]{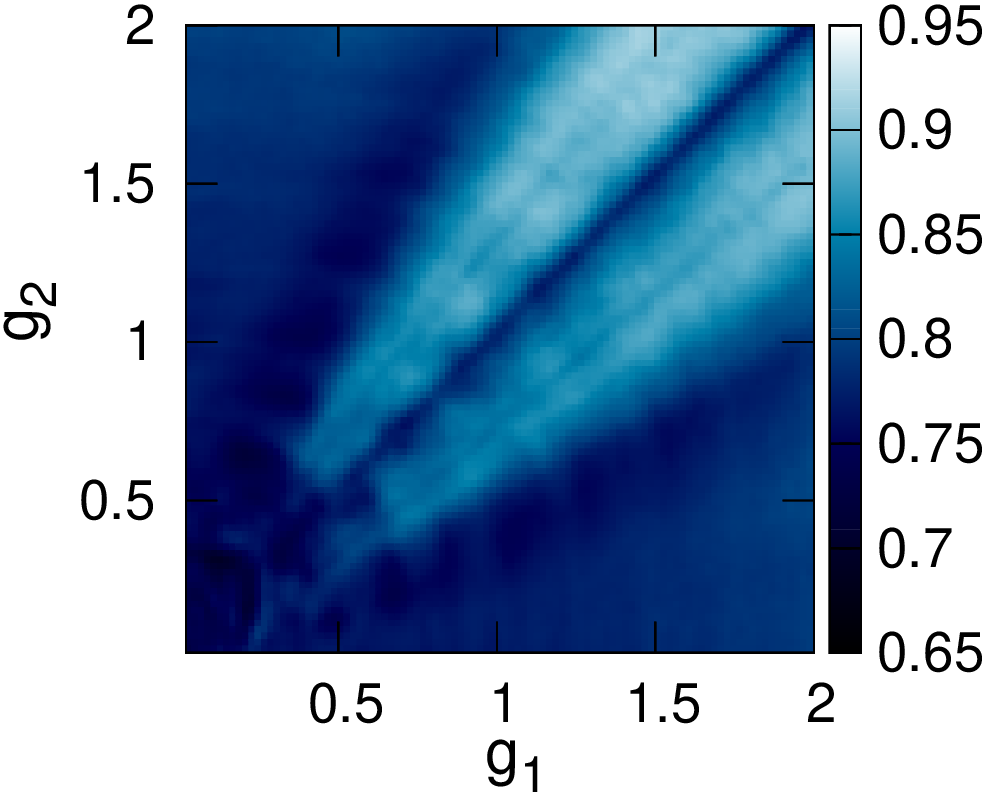}\label{fig:7v}}
    \subfigure[]{\includegraphics[width=0.195\textwidth]{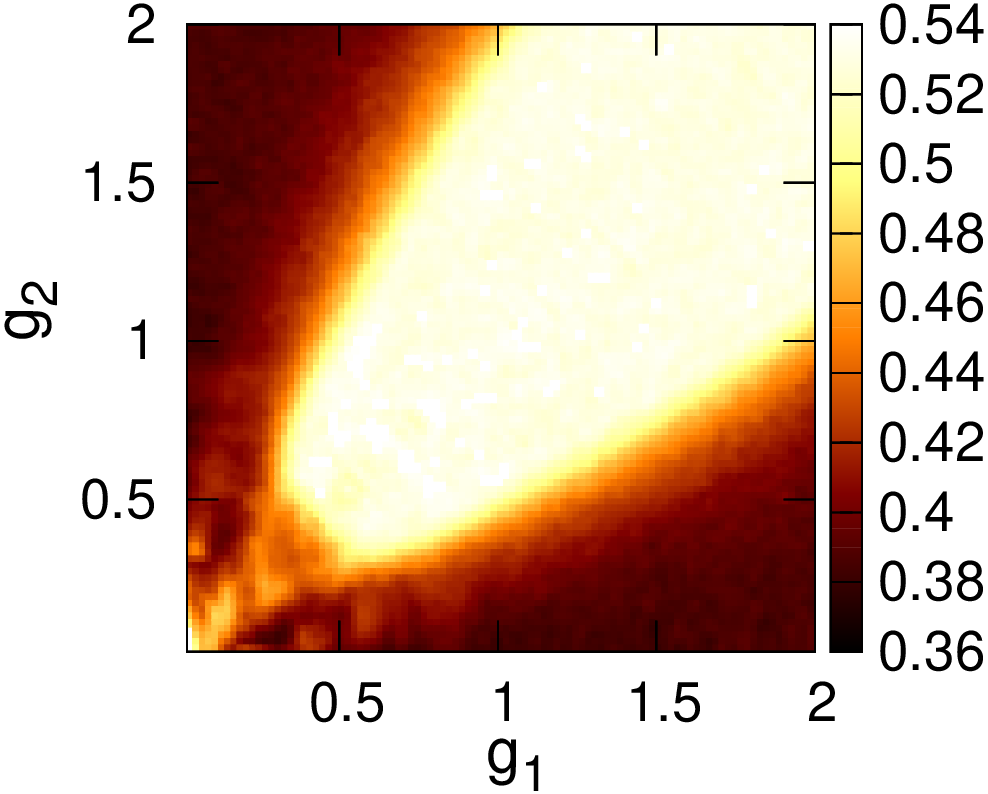}\label{fig:7w}}
	\subfigure[]{\includegraphics[width=0.19\textwidth]{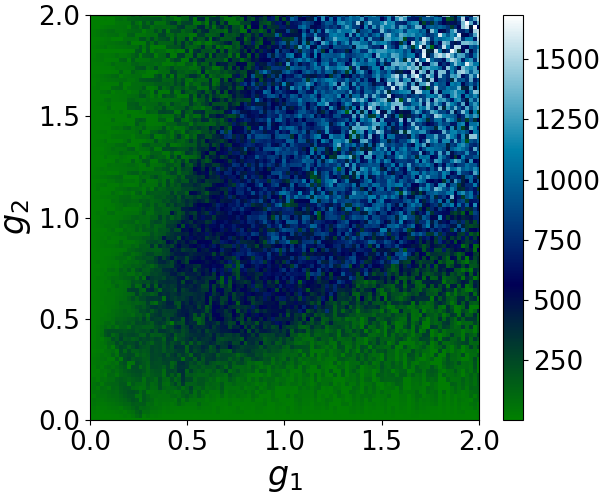}\label{fig:7x}}
	\subfigure[]{\includegraphics[width=0.19\textwidth]{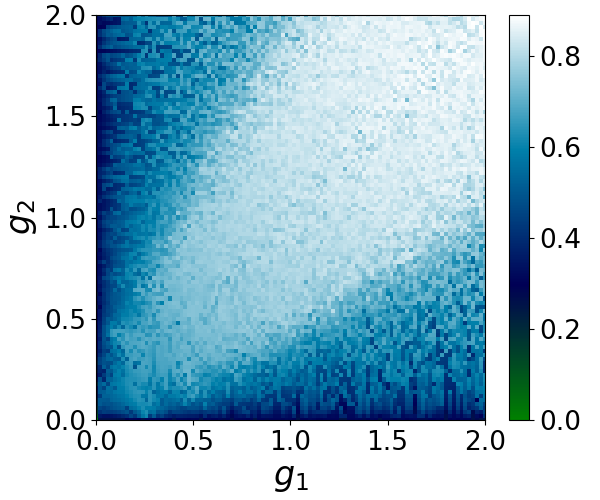}\label{fig:7y}}
	\caption{(a), (f), (k), (p), and (u): $\chi_{\text{lower}}$ [given
          in Eqn.~\ref{eqn:chi_lower}], (b), (g), (l), (q), and (v):
          $\chi_{\text{upper}}$ (given in Eqn.~\ref{eqn:chi_upper}),
          (c), (h), (m), (r), and (w): average consecutive level spacing
          ratio $\langle r\rangle$; (d), (i), (n), (s), and (x):
          participation ratio ($PR$) and (e), (j), (o), (t), and (y): the
          multifractal dimension $D_1$ for the middle excited state of
          the system for the resonance condition ($\omega=\omega_0=1$)
          considering $20$ spins and truncating the bosonic mode at
          gradually increasing values:
          $n_{\text{max}}=200,\ 300,\ 400,\ 500,\ 600$ (panels $1$ to
          $5$, respectively). }
	\label{fig:app2fig}
\end{figure*}     

    \section{Summary and conclusions}\label{sec_5}

    We first discuss the ground state phase transition from a normal
    phase to the super-radiant phase showing that a critical line
    $g_1+g_2=1$ separates the two phases. We see that the ground state
    energy density is almost constant in the NP whereas in the SP we
    find a broad range of energy densities, which are lower than that
    in the normal phase. The ground state number operator is almost
    zero in the NP while it is non-zero in the SP, indicating
    macroscopic excitations in the bosonic mode. By studying the
    participation ratio, we conclude that the ground state of the
    system exhibits multifractal features in the super-radiant phase
    with the participation ratio scaling as $PR_{\text{gs}}
    \propto\sqrt{N_D}$.

    Next, we explore the excited state features and find that the ADM
    also exhibits the excited state phase transition
    both for the resonant ($\omega=\omega_0$) and the off-resonant
    ($\omega\neq\omega_0$) cases. The ESQPT is nicely captured by the
    von Neumann entanglement entropy (between spins and bosons) as a
    function of eigenstate energies.  We observe that for the ADM
    there exist two cut-off energies separating the different phases:
    a lower cut-off energy (corresponding to the ground state energy
    along the line, $g_1+g_2=1$) and an upper cut-off energy
    (corresponding to the maximum energy at $g_1=g_2=0$ for finite
    $n_{\text{max}}$). Between these two cutoff energies, we find that
    the level statistics exhibits either Poisson statistics or
    Wigner-Dyson statistics depending on the values of the coupling
    parameters $g_1$ and $g_2$ suggesting a non-ergodic to ergodic
    phase transition. A study of the consecutive level spacing ratio
    of the system for the middle energy band (energy band between the
    lower and the upper cut-off energies) supports these findings.  It
    is convenient to introduce two new quantities (having the
    dimensions of energy) that correspond to the lower and upper
    cut-off energies of the spectrum in the super-radiant phase. The
    above energies are obtained by using the jumps in the VNEE as
    weights. These characteristic energies, which are a measure of the
    energies at which the corresponding von Neumman entropies of the
    eigenstates begin and end their plateau-like behavior, signal the
    ESQPT of this model. Remarkably when a phase diagram is obtained
    using these characteristic energies, we get a picture that looks
    very similar to the phase diagram obtained using level spacing
    ratios. Thus from our study, we conclude that
      the ESQPT and ENET are intimately related to each other for the
      anisotropic Dicke model ($g_1\neq g_2$). We checked that
      this connection is robust both for the resonant
      ($\omega=\omega_0$) and off-resonant ($\omega\neq\omega_0$)
      cases for the generic ADM; the diagonal direction ($g_1=g_2$)
      corresponds to the Dicke model exhibits special
      behavior~\cite{chavez2016classical}.  From a study of eigenstate 
      properties with the aid of participation ratio and multifractal 
      dimension $D_1$, we see that the normal-to-super-radiant phase 
      transition corresponds to the ground state undergoing a
      localized-to-multifractal transition.  On the other hand the
      ergodic-to-nonergodic transition corresponds to the middle
      excited state undergoing a delocalized-to-multifractal
      transition. The correspondence between the multifractal nature
      of the middle excited state and the nonergodic phase is also
      captured dynamically when we study participation ratio in a
      quench dynamical protocol.

    Finally, the ADM also exhibits yet another phase transition namely
    the temperature dependent phase transition. For $g_1+g_2>1$, there
    exists a critical temperature, $T_c$ above which the super-radiant
    phase disappears and the system comes back to the normal
    phase. Following the old work of Hioe~\cite{hioe1973phase} we
    write down an analytical expression for $T_c$ as a function of
    $g_1$, $g_2$. We show that the mutual information between two
    spins as a function of temperature proves handy to obtain an
    independent characterization of the thermal phase transition. A
    study of the mutual information suggests that for $g_1+g_2<1$, the
    system lies in the normal phase for all temperatures with a
    relatively lower value of mutual information, but for $g_1+ g_2>1$
    there exists a $T_c$ such that when $T<T_c$ the system lies in the
    SP with a relatively higher value of mutual information and for
    $T>T_c$ the system goes back to NP showing a temperature dependent
    phase transition.

\section*{Acknowledgments}
	We are grateful to the High Performance Computing(HPC)
        facility at IISER Bhopal, where large-scale calculations in
        this project were run. P.D. is grateful to IISERB for the PhD
        fellowship.  A.S acknowledges financial support from SERB via
        Grant No: CRG/2019/003447, and from DST via the
        DST-INSPIRE Faculty Award No. DST/INSPIRE/04/2014/002461.
        
	\twocolumngrid
	\appendix

	\section{Expression for the critical temperature for the TPT}\label{app_1}
        In this appendix we include a brief derivation of the
        expression for the critical temperature for TPT similar to the
        previous work~\cite{hioe1973phase}. We begin by recalling the
        expression of the partition function in the form of a double
        integral [Eqn.~\ref{eqn:partition_func}]:
        \begin{align}
          Z(N,T) =
          \int\frac{d^2\alpha}{\pi}e^{-\beta\vert\alpha\vert^2}\Big[
            \text{Tr}e^{-\beta h} \Big]^N\nonumber\\ =
          \int\frac{d^2\alpha}{\pi}e^{-\beta\vert\alpha\vert^2}\Big(
          2\cosh\Big[ \frac{\beta\epsilon}{2} \Big[ 1 +
              \frac{4(\lambda_1+\lambda_2)^2\alpha^2}{\epsilon^2 N}
              \Big]^{1/2} \Big] \Big)^N.
        \end{align}
        Rewriting the double integral using polar coordinates and defining  $y=\frac{r^2}{N}$ and
        \begin{align}
          \phi(y) = -\beta y + \ln\Big( 2\cosh\Big[ \frac{\beta\epsilon}{2}\Big[ 1 + \frac{4(\lambda_1+\lambda_2)^2 y}{\epsilon^2} \Big]^{1/2} \Big] \Big)
        \end{align}
        we can write:
        \begin{align}
          Z(N,T) = N \int_0^{\infty} d y\exp\Big( N \phi(y)  \Big).
        \end{align}
        Since we are interested in the thermodynamic limit where
        $N\to\infty$, we can invoke Laplace's
        method~\cite{jeffreys1999methods} to evaluate the integral as:
        \begin{align}
          Z(N,T) = N\frac{C}{\sqrt{N}}\max_{0\leq y\leq \infty}\exp\Big( N\Big[ \phi(y) \Big] \Big)
        \end{align}
        where $C$ is some constant. To find the maximum of the function $\phi(y)$, we compute its derivative:     
        \begin{align}
          \phi^{\prime} = -\beta + \frac{\beta(\lambda_1+\lambda_2)^2}{\epsilon}\frac{1}{\eta}\tanh\Big( \frac{\beta\epsilon\eta}{2} \Big)
        \end{align}
        where 
        \begin{align}
          \eta = \Big[ 1 + \frac{4(\lambda_1+\lambda_2)^2 y}{\epsilon^2} \Big]^{1/2}.
          \label{eqn:eta}
        \end{align}
        Putting  
        \begin{align}
          \phi^{\prime} = 0,
        \end{align}
        we have  
        \begin{align}
          \frac{\epsilon\eta}{(\lambda_1+\lambda_2)^2} = \tanh\Big( \frac{\beta\epsilon\eta}{2} \Big).
          \label{eq8}
        \end{align} 
        The hyperbolic tangent funtion is a monotonically increasing
        function and is bounded above by unity. Since $\eta \ge 1$ by
        definition [Eqn.~\ref{eqn:eta}], if
        $(\lambda_1+\lambda_2)^2<\epsilon$, there is no solution for
        Eq.~\ref{eq8}. On the other hand, for
        $(\lambda_1+\lambda_2)^2>\epsilon$, the solution depends on
        the value of $\beta$. The critical value of the inverse
        temperature $\beta_c$ can be computed by putting $\eta = 1$
        and is given by:
        \begin{align}
          \beta_c = \frac{2}{\epsilon}\tanh^{-1}\Big( \frac{\epsilon}{(\lambda_1+\lambda_2)^2} \Big).
        \end{align} 
        Substituting $\epsilon=\frac{\omega_0}{\omega}$ and 
        $\lambda_1=\frac{g_1}{\omega}$, $\lambda_2=\frac{g_2}{\omega}$, 
        we have an exact expression for the transition temperature:
        \begin{equation}
          T_c = \Big(\frac{\omega_0}{2\omega}\Big)\frac{1}{\tanh^{-1}\Big( \frac{\omega\omega_0}{(g_1+g_2)^2} \Big)}.
          \label{eqn:Tc_v2}
        \end{equation}

    \section{Robustness of our results against increasing bosonic cut-off}\label{app_2}
    In Fig.~\ref{fig:app2fig} we show the data corresponding to the
    ESQPT and ENET for $20$ spins and considering gradually increasing
    values of the bosonic truncation number: $n_{\text{max}} =
    200,\ 300,\ 400,\ 500, \ 600$ (panel $1$ to $5$ respectively) to
    check the robustness of our results. In each panel of this figure,
    we show $\chi_{\text{lower}}$ (the lower cut-off energy which is
    scaled by the minimum energy in the nornal phase),
    $\chi_{\text{upper}}$ (upper cut-off energy which is scaled by the
    maximum energy at $g_1=g_2=0$), consecutive level spacing ratio
    $\langle r \rangle$, participation ratio ($PR$) and the
    multifractal dimension ($D_1$) of the middle excited state
    respectively as a function of the coupling parameters $g_1$ and
    $g_2$. We notice that the results are qualitatively unchanged
    against increasing values of the bosonic truncation number
    $n_{\text{max}}$ and are also converging.

\twocolumngrid
\bibliography{refn2}
\end{document}